\documentstyle{article}

\begin{document}
\twocolumn[\hsize\textwidth\columnwidth\hsize\csname
@twocolumnfalse\endcsname
\date{ }
\title{The system with discrete interactions I:
Some comments about the principles of quantum theory 
}
\author{M. Yudin}
\maketitle
\begin{center}
{\em The Department of Mathematics and Statistics, 
The University of Melbourne, 
Parkville, Victoria,
Australia.}
\end{center}
\begin{abstract}
The model of the physical system with discrete interactions is based on the postulates that

(i) parameters of the physical system are defined in process of its interaction;

(ii) the process of interaction is discrete.

Consequently ordering of the events in the system is not automatically implied based on the values of time, but must be specified explicitly. This suggests the specific logic of events as well as complex character of the function, which describes stochastic behavior of the system, hence the parameters of the system are defined by Hermitian operators in the Hilbert space of functions.

The model gives an essential way to introduce the basic notions and obtain the results of the quantum theory, to derive the meaning of the main physical parameters, such as momentum or electric charge.
Based on the proposed postulates, Schr\"{o}dinger equation for the particle is deduced in a way similar to inferring of Smoluchowski equation in the classical statistical mechanics.

Similarly, the equations of motion of linear fields are established by considering Smoluchowski-type equations for amplitudes of harmonics. 

The concept of discrete interaction gives a way to de-couple the object of experiment and a testing device (the traditional source of controversy in quantum theory), though at expense of limitation of the principle of locality, which is considered to follow from the discrete nature of interaction. 
\end{abstract}

$\;\;\;\;\;\;\;\;\;$ {\bf PACS} 03.65.bz, 03.67.-a
\vskip2pc]

\large
\section{\rm Quantum theory: positivism versus realism}
The first question to be asked, when addressing the problem of interpretation of quantum theory, may sound, why does the quantum theory need some kind of special interpretation while the classical does not. The pragmatic approach, put forward by Copenhagen school and accepted as the foundation of the theory, is simply to postulate that the conceptual framework expressed as mathematical formalism and used in the theory is the most fundamental and does not need interpretation using the traditional set of metaphysical concepts. 

It is commonly accepted that quantum theory is complete: the formalism of the theory gives adequate mathematical description of all relevant experimental data. Arguing that only the body of experimentally visible phenomena constitutes the meaningful concept of the nature, the conclusion may be drawn that as long as the superimposed models of reality do not have testable consequences different from those predicted by the current mathematical formalism, the metaphysical appendages must be ignored. One can remark that due to fundamental to the theory principle of uncertainty, the attempts to assume a fine structure behind the phenomena may not lead to any verifiable deviation of experimental results from already predicted. That means, the theory protects itself from establishing of any kind of underlying metaphysics, which even hypothetically may be compatible with the classical.
On the other hand, one may argue that as far as human knowledge is concerned, the meaningful concept of understanding would imply representation of the phenomena using the predefined (in Kantian sense) language of concepts attributed to cognition.
In this respect, as long as the theory denies the attempts to perform this translation, it states that the nature, at least in some of its manifestations, {\em can not} be understood, but simply recorded. 
This reduces the goal of science to ``augment and order of our experience" \cite{bohr1}, rather than to establish its conceptual blueprint using the predefined metaphysical framework.

The radical expression of the position formulated by Stapp \cite{stapp} may sound, rather than being the description of the real essence of the phenomena, quantum theory is ``a procedure by which scientists predict probabilities that measurements of specified kinds will yield results of specified kinds in situations of specified kinds". 
Using the term procedure instead of theory does not mean that it completely denies metaphysics (no theory can operate if the concepts of space or time are omitted), this only articulates the fundamental differences between two independent conceptual frameworks defined for quantum, positivistic, or more specifically pragmatic, and classical, arguably realistic theories, without making an attempt to reconcile them.
 
The argument to justify this pragmatic approach, put forward by Bohr may sound that unlike at the macroscopic level, the language of the nature at the microscopic level is incompatible with conceptual language of cognition, so that only the formal translation seems to be possible. That is, the only meaningful language to convey information when describing the phenomena, would to specify the input and expected output of possible experiments. The results of experiment may be estimated based on mathematical formalism, established without any kind of conceptual constraints. Referring to SQUID experiments \cite{squid}, \cite{leget} (which give an example of non-classical macroscopic behavior), 
we must acknowledge the presence of two independent conceptual frameworks (observable parameters versus operators in Hilbert space) at the same macroscopic level.

It may be noted that there is no other link between mathematical formulae and description of an experiment rather then the language of metaphysical concepts, as well as visualization.
To be able to draw the relation between the formulae and the set of dots on the plate in the case of Lennard-Jones experiment, one has to think in terms of wave, that is to visualize the behavior of quantum particle in space. 
To be described and interpreted, the experiment must be understood; for each particular case the consistent model of underlying process still has to be drafted. The only difference with the realistic approach would be that these models, being presented in terms of conventional concepts, for the different experiments conducted on the same substance, are seemingly incompatible with each other. 
Interpretation, which in the case of realistic approach would be associated with the substance (electron -- a particle, light -- a wave, with all conventional properties to follow), needs to accompany the particular experiment. The condition that seemingly incompatible concepts should be attributed to the same substance requires separate models.\footnote{The concept of a particle as an object located in space and the concept of motion of the medium distributed in space (wave), are associated with the same substance -- quantum particle, for each moment of time}. Instead of a single self-consistent picture, we have incompatible fragments, with the particular one to be drawn when appropriate (defined by conditions of an experiment). That implies strong coupling of the object and the subject of experiment, with the predominant belief that the subject of experiment creates the state of the object, rather than discovers it: fundamentally positivistic point of view.

The positivistic character of the theory is more hidden in the contemporary approaches, which claim that mathematics being the fundamental language of the nature should as well be used as the language of its understanding. That is, mathematical definitions are interpreted in a manner normally reserved for metaphysical concepts: the mathematical formalism is regarded as the most fundamental conceptual description of the nature which in this case has to be established {\em a priory} in order to be able to comprehend the phenomena indicated in experiment. No mental picture is possible, unless derived from mathematical model\footnote{For example the quantum field defined as a set of operators in the functional Hilbert space implies the particle-like properties of the field, which in effect is considered to be the consequence of this generic statement (not vice a versa). In this case the mathematical definitions of operators and Hilbert space are effectively promoted to the rank of metaphysical concepts, so that as many {\em new} metaphysical concepts are introduced as needed to formulate the axioms.}. To understand the nature would in this case mean to establish the set of definitions and formulae, so that a mathematical model can be perceived as a picture of underlying reality. We obviously changed the meaning of the word understand to what used to be called "present in mathematical terms", rather then "express in terms of predefined conceptual language".
\\
\\
\small
Formalization of Copenhagen interpretation, basically completed by von Neumann \cite{v_neumann} evolved to extensive development of axiomatic approaches within the theory \cite{bunge,zierler,piron,gerlich}.
The alternative approach would to examine the logical aspects of quantum behavior. The subject addressed by Birkhoff and von Neumann \cite{birkhoff} as early as in the thirties is constantly attracting much of attention \cite{mackey, beltrametti, ptak,d_chiara, BrigesSvozil}.
In either case, the theory confirms that existence of two distinct conceptual or logical frameworks is necessary for description of the physical nature (as long as the quantum theory is fundamentally based on the principle of correspondence\footnote{The classical system as such may not necessarily be postulated, but established as a limiting case of quantum in the environment with decoherence, still the equations of the quantum theory (mechanics and field theory) are derived based on the classical, which, unless quantum equations are established from the first principles, must be defined independently.}, the classical theory cannot be interpreted as a mere limiting case of quantum, but an independent theory of its own). 

For completeness, we must also mention realistic models developed to describe quantum phenomena. 

We must refer to historically significant model put forward by De Broglie \cite{debroglie}, in which he describes the behavior of a quantum particle as a kind of a classical one, with the wave function interpreted as an internal vibration of the particle (the wave function is interpreted similarly to the field). This was generalized by D. Bohm in his model of quantum behavior \cite{bohm} where a set of hidden parameters with the classical properties is introduced. Since then, the approaches relying on hidden parameters or realistic (field) interpretation of the wave function are constantly attracting certain amount of attention, to mention just a few publications \cite{dewdney,holland}.

Here however we need to remark, that due to limitations of their scope mainly to wave-particle aspect of motion (no quantum statistics or field theory has been properly developed), though primarily due to the fact that interpretation of the wave function as the field, rather than the probability leads to numerous problems when consider measurement (instant collapse of the wave function in this case would violate the basic principle of special relativity), realistic models were never able seriously to challenge the mainstream interpretation.
\\
\\
\\
\large
Not reflecting on the fact that "quantum" and "classical" languages should coexist (the language of metaphysics still used in all areas of human knowledge cannot be discarded), we remark that the theory still cannot be called realistic in the plain meaning of the word. This stems from acknowledgement that in this case the universal description of the nature is the wave function with the meaning of probability. Unlike classical physics, when the probability is basically a measure of our lack of knowledge\footnote{If we say that a molecule of gas is located in the given area with a certain probability, we assume that the molecule is definitely located somewhere, but our guess about the precise location would have only a chance of success}, in the quantum theory the wave function is the most fundamental description of the nature. So that we specify the wave function of the system without an assumption that the (value of) underlying property exists at all in the current environment, moreover in some circumstances, the existence of the parameter would be denied in principle. When describing electron by the wave function $e^{i({\bf k}{\bf r} - \omega t)}$ in the double slit experiment, we do not assume that the electron has any particular single coordinate ${\bf r}$. In short, unless some ever-present hidden variables are introduced, we prescribe the probability to the parameter, which does not exist in the particular moment in time. It will exist, provided the appropriate observation is made, but in this case we returned to the positivistic approach: the reality cannot be imagined independent of its perception. The theory cannot be called realistic as long as the meaning of the world real is not changed from actually existing (independent of our actions), to potentially possible, which would exist in this or that way depending on conditions of experiment. In this respect, the quantum theory is the only scientific theory, which needs to question the concept of reality as such.

The interesting aspect to note is that, though any philosophical system can be imposed, the classical theory is essentially realistic. A single coherent metaphysical description is associated with the substance. This description is not coupled with the process of observation and by virtue of its uniqueness, can be regarded as conceptual model of underlying reality, rather than depend on the experiment, involving the substance. The reality can be understood without the {\em necessity} to refer to experiment in the conceptual model. On the other hand, at the quantum level, the nature seems to present itself in essentially positivistic fashion: unless resorting to unverifiable speculations, no model of reality can be established without specific reference to the experimental environment. The formalism of quantum theory, can be transformed to the classical in the case when $\hbar\rightarrow 0$. This means that fundamentally positivistic model can be transformed to realistic by modifying the parameter in its mathematical description. This could be fuelling the search of underlying conceptual system to be used as a unified approach.

In the present work the author considers the model of behavior of the system with {\em discrete interactions}. It is demonstrated that the description of the system obtained within the present approach coincides with the description established using the conventional formalism of quantum theory, though we start from supposedly realistic conceptual model rather than mathematics. 
The material is presented in a way that the set of axioms, as conventionally defined for quantum mechanics, is reconstructed. 
This may be interpreted as an indication that the specific features of the behavior of the quantum system can be attributed to fundamentally discrete character of interactions.

\section{\rm Discrete interaction: Parameters of the system}
We discuss how the behavior of the physical system will be apparently changed if the system is considered within the model of {\em discrete interaction} (DI), compared with the classical model with the continuous interaction. 

To continue we clarify what is meant here by discrete interaction. When consider interaction of two systems as specified within quantum theory of measurement, interaction within the system is presented by perturbation term in Sch\"{o}dinger equation, defined by potential acting within short though continuos interval of time (the approximate interpretation of potential as a classical field is used).
On the other hand, at the fundamental level interactions are supposed to be carried by fields, which are presented as sets of quanta. Interaction with a quantum of the field is described as a collision, taking place in a discrete moment of time. Anywhere in the text, though considering non-relativistic model, we will be using the term discrete interaction in its fundamental meaning, while interpreting potential term in Sch\"{o}dinger equation not as an indicator of continuous interaction, but a function which describes inhomogeneous properties of space. 
Rather than investigating the detailed picture, which is the subject of theory of measurement, we add an assumption that interaction between the classical system and the DI-system used in the current model is also discrete.

As an example we consider non-relativistic motion of the particle in space (so we can avoid discussion of the problems related to measuring of time which will be done later).
\\
\\
{\em Physical parameters and measurement.}
\\

The particle (material point) is a physical system the state of which is characterized by the set of parameters, such as coordinates and momentum. The particular values of the 
\\
parameters specify the different states of the particle.\footnote{We do not discuss the parameters, which characterize the particle as an object, and do not relate to its particular states, such as mass.}
In speaking about the coordinates, we should bear in mind that their particular values have very limited physical meaning unless the particle will interact with the external system, directly or indirectly, so that its position can be detected and identified.
This would apply to any parameter of a physical system.

Consider the simplest case of locating of a particle in space.
In the classical model, when we say that the particle has the particular position in space, we mean that if at a given moment of time we locate a detector in the given point, the particle will interact with it. If we are able to locate detectors in any position in space, the particle will continuously interact with them in the different positions, so we say that the particle is moving in space. It may be noted, that the position (as specified by the value of coordinate) may be prescribed to the particle in the similar fashion, whether the particle is interacting with a real detector or will interact, would the detector be located in the particular point. In the classical system the position may be always prescribed to the particle because, at least potentially, the particle may {\em continuously} interact with the detector.\footnote{Consequently, in the case of the classical, deterministic system, the measurement have quite a restricted area of application: the values of parameters, estimated according to the initial conditions, must coincide with the measured values, so that the only one, though infinitely accurate measurement is needed.} 

In the case of discrete interaction, by definition, at the particular moment of time the particle may not interact with any detector located in any point in space. On the other hand the particular position may be prescribed to the particle only if been detected. If the particle can interact with the detector only in discrete moments of time, {\em the particular values of the coordinate may be prescribed to the particle only at these moments of time}. 
This leads to the question how to describe the state of the particle in between interactions, which becomes the central task of the model. The most general way is formally to prescribe to the particle {\em all possible positions} (a set) with the constraint that the particle should be able to pass from these positions to the detected (observed) position at the moment of detection (observation).

The model allows several rules to define the set of positions between detection, or more generally, between interactions.

(i) An empty set: no position is prescribed to the particle in between interactions. If this applies to all measurable parameters, the model does not specify the behavior of the system between interactions, unless some hidden parameters are introduced;

(ii) A unique position (a single value of the coordinate) is prescribed. If the particular value of the coordinate is defined according to the principle of the least action the model is similar to the classical: the discrete character of interaction cannot be manifested;

(iii) A set, which may include more than one position, is prescribed. This requires a nontrivial model of the behavior of the system in between interactions. 
We consider this, more general type of behavior. 

The other question to be addressed is how the different parameters relate to each other. 
We define the parameter as a set of values, which identify the observable state of the system. The observable states of the system are described by the different single values of the parameter (e.g. the different positions of the particle are specified by the different values of its coordinate).
The model with discrete interactions permits the situation when not all parameters of the DI-particle are defined simultaneously.
Indeed, the particular parameter of the particle can be detected due to interaction of the particular type, which may imply specific behavior of the particle. In order to detect the position, the particle should interact with a detector in the particular point in space, and in order to measure the momentum the detector should allow the particle to pass between the neighboring positions. The process of interaction when the particle enters a given detected state means certain type of behavior of the particle itself, which may be triggered by the detector. The condition that the parameter may be detected only in the particular discrete moments in time means that the process, which allows the detector to detect the parameter, can be triggered in discrete moments in time. The particular types of behavior of the particle, which allow us to detect different parameters, may happen to be incompatible with each other. In the classical case the problem of compatibility does not appear due to the continuous character of interactions: the values of all parameters are detected continuously so they must be compatible {\em ad definicio}.

To discuss these questions in detail, we give formal definitions of the principles fundamental to the model and consider the main results, which follow from these principles.
 
\section{\rm The basic principles: Formal definition}

We present our statements in the general form applicable to an arbitrary physical system and promote them to the rank of the postulates, which we formulate as follows.
\begin{itemize}
\item {\em The particular values of parameters are defined in process of interaction}, which is the specific kind of behavior of the system, possible in the specific environment (in particular created by the detector). 

\item {\em The process of interaction is discrete.} Consequently, between interactions the particular values of parameters are not defined, the set of values (not necessarily one) may be prescribed to each parameter of the system.
\end{itemize}

We assume that the states of the system between interactions can be derived based on the states of the system defined by interaction\footnote{the alternative would mean hidden behavior of the system}; these are the states defined by single values of the particular parameter. 
\\

{\em Note}[$\ddagger$]:  The postulates imply existence of the process due to which the single unique values of the parameter are identified (interaction of the particular kind). 
\\

{\em Definition:} The state which corresponds to a single unique value of the given parameter we call an elementary state in the given representation \footnote{Analogue of the eigenstate (pure state) in the quantum mechanics.} (all the states are elementary in the system with the continuous interaction). 
\\

We identify elementary states of the system based on the type and values of the physical parameter, which the system currently possesses. In particular, for the particle an elementary state in the coordinate representation
($R$) is the state for which the position of the particle is defined by a single coordinate ${\bf r}_{0}$. An elementary state in the momentum representation ($P$), is the state in which the particle has the given value of momentum ${\bf p}_{0}$; we assume that in the coordinate representation this state will be defined by the appropriate set of coordinates.

In general, the set of states $\{{\cal A}_{i}\}$ which may be prescribed to the system in between interactions is defined according to the type of possible interaction with external system (detector of the given type which detects the parameter $a$). The states ${\cal A}_{i}$ (subsets of the set $\{{\cal A}_{j}\}$), which correspond to single different values of the physical parameter $a$ we call elementary states in the representation ${\cal A}$.
We assume that two particular representations may be related, that is, an elementary state in one representation will be presented by a set of elementary states in the other representation.

The values of parameters of the DI-system are not specified in between interactions. This means that between interactions {\em all possible values of the given parameter (not necessarily one)} may be prescribed to the system: the system may possess all the values of the parameter, which are {\em allowed according to possible constraints}.
The state of the system between interactions can be described as a set of elementary states (which correspond to the particular values of the parameter). This set of states is formally defined according to the following constraints: it includes all the states, to which the system can pass from the initial defined state during the time since the moment of interaction, or alternatively, from which it can pass to the next defined state during the time since the current moment till the moment of the next interaction.
The system is described by the set of elementary states $\{A_{i}\}$, which defines its state between interactions. Note, that applying this principle for the description of a particle in space, we conclude that it is essentially non-local (a set of coordinates is prescribed to the particle at the same time). It becomes local only at the moment of interaction of specific kind (detection of position), which may be understood as something similar to choosing of the particular coordinate from the set. 
We assume that for the particular moment of time in between interactions, the set of states is prescribed to the system; that means no ordering for the states is defined.
\\

{\em Definition:} The state of the DI-system between interactions defined as a set of elementary states, which are entered by the DI-system at the particular moment of time between interactions, we call a non-ordered state.
\\

The similar would be true for each particular moment at the interval of time between interactions (the set of states is prescribed to the system for each moment of time at the interval). 
If the same non-ordered states (the same set of elementary states) are prescribed to the system at each moment at the interval of time, we can say that this non-ordered state is prescribed to the system at this interval of time.
This in particular implies that {\em ordering} of elementary states, which belong to the non-ordered state, {\em is not defined during this interval of time}. Two different elementary states, prescribed to the DI-system in the subsequent moments of time, do not necessarily imply transition from one of these states to the other. The states may belong to the same non-ordered state, which is prescribed to the system at these moments of time, thus the transitions between the states should be specified explicitly.    

In the classical system we do not need specially to consider the states entered non simultaneously with the given one. Two states described by the different values of the same parameter cannot be entered simultaneously and always relate to subsequent moments of time. On the other hand for the DI-system even if the states of the system are defined in the subsequent moments of time, they still may be considered as simultaneous in the sense that no ordering between the states is defined. Ordering of the states is not automatically implied by their sequence in time; the set of states into which the system passes from the current state, or alternatively from which the system passes into the current state, should be explicitly prescribed to the system at each particular moment of time. 

To take into account this aspect of behavior of the DI-system, we extend the definition of non-ordered state. 
\\
 
{\em Definition:} The state of the DI-system between interactions defined as a set of elementary states, such that for each particular moment of time the system stays at the states of the set (non-ordered state), or passes into (from) the states of the set (ordered state) we call an intermediate state.
\\
    
Intermediate state includes the particular non-ordered state defined for the given moment of time and the set of states into which the system passes from the given non-ordered state, or from which the system passes into the given non-ordered state.

\section{\rm Behavior of a system between interactions}

\subsection{\rm Logical Model}
We continue with the example of motion of the particle in space taking as a starting point the postulates formulated above.
According to these postulates, each parameter of the particle (coordinate, momentum) is detected and identified as a result of an interaction of the appropriate type. 

Consider two elementary states attended by the DI-particle at the given moment of time between interactions, say two positions of the particle. These positions may belong to the same non-ordered state, in this case ordering for them is not defined; the positions are described as if the particle simultaneously enters both of them.
Alternatively the particle passes between the positions, in this case certain formal ordering can be defined for these positions: the positions are described as if the particle enters them in a sequence, non simultaneously.

Generally, for two states ($1; 2$), which are prescribed to the system at the particular moment of time, the formal ordering can be defined in the sense that these states do not belong to the same non-ordered state (the system passes from the state $1$ to the state $2$). We assume that 
\\

{\em Definition}[$\dagger$]: The particular ordering relation (ordering, non-ordering) of the given state $\cal A$ and an elementary state which belongs to a non-ordered state implies similar ordering relation between $\cal A$ and each elementary state of this non-ordered state.
\\
\\
Based on [$\dagger$] we can define the relations for an arbitrary set of elementary states 1, 2 and 3 which belong to an intermediate state as follows.
\\

{\em Proposition}: [$\dagger$ $\dagger$]: ordering defined for the states 
1, 2 and no ordering defined for 2, 3 implies ordering for 1, 3.\footnote{The proof follows from the definition.}
\\

Note that [$\dagger$ $\dagger$] also means that:
\\

{\em Proposition}: 

(i) no ordering for 1, 2 and no ordering for 2, 3 implies no ordering for 1, 3;

(ii) ordering for 1, 2 and ordering for 2, 3 implies no ordering for 1, 3.\footnote{In both cases the alternative will violate [$\dagger$ $\dagger$].}
\\

The concept of ordering (no-ordering) of 
\\
states can be interpreted as a logical relation defined for the states, which can be used without explicit reference to the particular non-ordered state.
According to the definition, for any two different elementary states, which belong to the same intermediate state, two types of the behavior can be defined:

1. the system passes from one state to the other (event ${\cal V}_{1\wedge 2}\equiv {\cal V}_{1,2}$): formal ordering is defined for the states;

2. the system stays in the state, which incorporates these two states (event ${\cal U}_{1 \wedge 2} \equiv {\cal U}_{1,2}$): no ordering is defined for the states.
\\
This may be considered also as a definition of two types of events: entering of two non ordered states ${\cal U}$, transition between the states ${\cal V}$ for the given intermediate state. 
\\

{\em Definition} The states $a$, $b$ are logically independent, if they are defined to be mutually exclusive: the states cannot be entered simultaneously no transition between the states is possible.

The states $a$, $b$ are logically related if they can be entered simultaneously, or transition between the states is defined. 
\\

The definition of events ${\cal U}_{i\wedge j},{\cal V}_{i\wedge j}$ for the system with two logically related states and the property  [$\dagger$ $\dagger$] implies the following rules of the behavior for the system  with three logically related states  
\begin{equation}
{\cal V}_{1\wedge 2\wedge 3}=({\cal V}_{1\wedge 2}\wedge{\cal U}_{2\wedge 3})
\vee({\cal U}_{1\wedge 2}\wedge{\cal V}_{2\wedge 3}),
\label{vv}
\end{equation}
The system passes from the state which includes 1 into the state which includes 3 (ordering for 1 and 3 is defined) via the state 2, if the system passes from the state which includes 1 into the state which includes 2, provided the state which includes 2 also contains 3, or the system passes from the state, which includes 1 and 2 into the state which includes 3 (fig.1).
\begin{equation}
{\cal U}_{1\wedge 2\wedge 3}=({\cal U}_{1\wedge 2}\wedge{\cal U}_{2\wedge 3})
\vee({\cal V}_{2\wedge 1}\wedge{\cal V}_{2\wedge 3}),
\label{uu}
\end{equation}
The state of the system includes 1 and 3 (ordering for 1 and 3 is not defined), provided this state also includes 2, or the system passes to 1, 3 from 2, if the state of the system includes both pairs of states 1, 2 and 2, 3 or the system passes from 2 to 1 and from 2 to 3.  
These rules can be generalized by induction to describe the case of any number of logically related states.

For transition to an arbitrary state 2 from any of the two independent states $a$, $b$ we have (for mutually excluding events $\vee$ has the same meaning as $\oplus$)
\begin{equation}
{\cal V}_{(a\vee b)\wedge 2}={\cal V}_{a\wedge 2}\vee {\cal V}_{b\wedge 2},
\label{v3}
\end{equation}
the system passes from $a$ to 2, or the system passes from $b$ to 2;
\begin{equation}
{\cal U}_{(a\vee b)\wedge 2}={\cal U}_{a\wedge 2}\vee {\cal U}_{b\wedge 2},
\label{v4}
\end{equation}
the state of the system simultaneously includes $a$ and 2, or $b$ and 2.
This is similar to the relations of the classical theory.
\\

{\em Definition}: 
A transition from the state $\Psi_{1}$ to the state $\Psi_{2}$ via all possible elementary states $x_{k}$, assumed all transitions are taking place independently, which is described as
\begin{eqnarray}
{\cal V}_{\Psi_{1}\wedge \Psi_{2}}={\cal V}_{\Psi_{1} 
\wedge x_{0}
\wedge \Psi_{2}}\dots\vee
\dots{\cal V}_{\Psi_{1}
\wedge x_{k}
\wedge \Psi_{2}}\dots,
\nonumber\\
{\cal U}_{\Psi_{1}\wedge \Psi_{2}}=
{\cal U}_{\Psi_{1}
\wedge x_{0}
\wedge \Psi_{2} }\dots\vee
\dots{\cal U}_{\Psi_{1}
\wedge x_{k}
\wedge \Psi_{2}}\dots,
\label{uu1}
\end{eqnarray}
we call unconditional transition (or simply transition) between the states $\Psi_{1}, \Psi_{2}$\footnote{Note, for elementary states, unconditional transitions between the different states are not allowed. The contrary would violate the definition of an elementary state as the state, which corresponds to the one and the only one value of the parameter.}
\\

Consider a particular state $\Psi_{p}$ and a set of elementary states $X$.
For each elementary state $x\in X$ we can define events of the type
${\cal V}_{\Psi_{p}\wedge x}$, ${\cal U}_{\Psi_{p}\wedge x}$. This may be considered both as the definition of $\Psi_{p}$ in terms of elementary states $x$ and the 
process of transition from $\Psi_{p}$ into each of the states $x$.
The case of the special interest is when the function $\Psi_{p}(x)$ identifies an elementary state $p$ in representation $P$. Assume the set of functions $\Psi_{p}(x)$ is defined for each $p\in P$. In this case we say that the representations are related in the meaning that each state $p$ in the representation $P$ can be presented in terms of set of states $x$ of the representation $X$.  

\large
\subsection{\rm Probabilistic model}
{\em Probability and events.}

As long as we prescribe to the system a set of elementary states at the same time, its behavior becomes essentially uncertain in the meaning that not the particular one, but any state, which belongs to the set, may be detected.
In the classical case, uncertain behavior of a system is described by the means of the probability theory. Generally, the probability theory may be used for the description of a physical system if:

1. In the case of a test ${\cal A}$ with possible unrelated results ${\cal A}_{a}$, ${\cal A}_{b}$, for the events ${\cal A}_{a \vee b}$, ${\cal A}_{\neg{a}}$, algebra of probabilities corresponds to logic of events according to the formulae 
\begin{equation}
Pr({\cal A}_{a \vee b})=  Pr({\cal A}_{a})+Pr({\cal A}_{b}),
\label{prob_0}
\end{equation}
\begin{equation}
Pr({\cal A}_{\neg{a}})= 1 - Pr({\cal A}_{a})\geq 0, 
\label{prob1}
\end{equation}
and for two independent tests 
\begin{equation}
Pr({\cal A}_{a\wedge b})= Pr({\cal A}_{a})Pr({\cal B}_{b}),
\label{prob0}
\end{equation}
this can be generalized for any set of events based on the rules of standard propositional logic;

2. Relations between frequences of events in a physical system correspond to the relations between the appropriate probabilities.

It may be noted that the fundamental relation between frequency of events and the value of probability prescribed to the event follows from the law of large numbers.
The law of large numbers can be derived for a set of independent events, such that

(i) the set is complete (may be extended to the complete set): all possible results of testing are represented by the events, which belong to the set;

(ii) occurrence of one of events in the testing, implies that other events do not take place in the same testing
{\em (Exact testing)}.

In the case of exact testing proportionality between the frequency of events and the probability, defined as a measure in the space of events (with algebraic properties (\ref{prob_0}-\ref{prob0})) may be established (see for example \cite{jeff}). 
\\
\\
{\em Generalized probability.}
\\

We define numerical values $U_{1,2}\equiv U({\cal A}_{1\wedge 2})$, 
$V_{1,2}\equiv V({\cal A}_{1\wedge 2})\in R$; $|U_{1,2}|<1$, $|V_{1,2}|<1$, so that the value $V_{1,2}$ is prescribed to the event ${\cal V}_{1\wedge 2}$ and the value $U_{1,2}$ is prescribed to the event ${\cal U}_{1\wedge 2}$, as specified for ${\cal A}_{1\wedge 2} = \{ {\cal U}_{1\wedge 2}, {\cal V}_{1\wedge 2}\}$ Addition and multiplication of $U_{i,j},V_{i,j}$ is defined according to (\ref{prob_0}), (\ref{prob0}), based on relations between the events 
${\cal U}_{i,j}, {\cal V}_{i,j}$.
For $V_{1,2}$ (similarly to intensity of the flux) we define $V_{1,2}=-V_{2,1}$.

We define the algebra for $U_{i,j},V_{i,j}$, so that it relates to the logic of events according to the rules similar to these used in the standard probability theory (\ref{prob_0}), (\ref{prob0}).
That is, for transition to the state 2 from two logically unrelated states $a$, $b$, according
to (\ref{prob_0}), we have
\begin{eqnarray}
V({\cal A}_{(a\vee b)\wedge 2})=V({\cal A}_{a\wedge 2})+
V({\cal A}_{b \wedge 2}),\nonumber\\
U({\cal A}_{(a\vee b)\wedge 2})=U({\cal A}_{a\wedge 2})+
U({\cal A}_{b \wedge 2}).
\label{gpv1}
\end{eqnarray}
Generally, recalling the formulae (\ref{vv}),(\ref{uu}), we can define $U_{i,j},V_{i,j}$ values in the system with the three logically related states (which describe transitions between the states 1, 3 via the state 2) based on the  values $U_{i,j},V_{i,j}$ for the systems with the two states (recalling $V_{1,2}=-V_{2,1}$)
\begin{eqnarray}
V(&{\cal A}&_{1\wedge 2\wedge 3}\;) \nonumber\\ 
&=& 
V({\cal A}_{1\wedge 2})U({\cal A}_{2\wedge 3})+
U({\cal A}_{1\wedge 2})V({\cal A}_{2\wedge 3})\nonumber\\
& \equiv & 
V_{1,2}U_{2,3}+U_{1,2}V_{2,3},\nonumber\\
U(&{\cal A}&_{1\wedge 2\wedge 3}\;) \nonumber\\ 
&=&
U({\cal A}_{1\wedge 2})U({\cal A}_{2\wedge 3})-
V({\cal A}_{1\wedge 2})V({\cal A}_{2\wedge 3})\nonumber\\
& \equiv & 
U_{1,2}U_{2,3}-V_{1,2}V_{2,3}.
\label{gpv}
\end{eqnarray}

That is, the algebraic relations defined for $V_{i,j}$, $U_{i,j}$ correspond to the algebraic relations defined for real and imaginary part of the complex number $c({\cal A}_{j \wedge i})=<j|i>$, so in  the abbreviated notations (\ref{gpv1} - \ref{gpv}) can be presented as 
\begin{equation}
<3|1>=<3|2><2|1>,
\end{equation}
\begin{equation}
<2|a\vee b>=<2|a>+<2|b>,
\end{equation}
$U_{i,j}=Re<j|i>$, and $V_{i,j}=Im<j|i>$.

Consider a particle, which may enter the states with the different positions $x$. In terms of introduced characteristics of transitions we define a complex function $\Psi(x)\equiv <\Psi|x>$, which is prescribed to the given intermediate state of the particle $\Psi$ provided this state includes the position $x$ ($Re \Psi(x)=U(x)$), or that the particle passes to the given state $\Psi$ from the state, which includes the position $x$ 
($Im \Psi(x)=V(x)$), $\forall x \in R$.

Generally, for an elementary state $a_{i}\in A$ the definition of function $\Psi(a_{i})$ for the the event $<\Psi|a_{i}>$ implies that the event $<a_{i}|\Psi>$ is described by complex conjugate of $\Psi(a_{i})$. We can specify scaling of the function $\Psi(a_{i})$ according to the formula
\begin{equation}
\sum_{A} \Psi(a_{i})\Psi^{*}(a_{i}) = 1,
\label{scal_1}
\end{equation}
and similarly for the continuous set of states.

For the discrete set of states consider the density matrix, as defined in the quantum theory
$P_{i,j}=\Psi_{b}(a_i)\Psi_{b}^{*}(a_j)$. 
The special case when the density matrix is diagonal 
\begin{equation}
P_{i,j}=<a_{i}|b><b|a_{j}>=p_{i}\delta_{i,j},
\label{prob}
\end{equation}
or more generally
\begin{equation}
U_{a_{i},a_{j}}=Re(<a_{i}|\Psi><\Psi|a_{j}>) = u_{i}\delta_{i,j}.
\end{equation}
is the condition of exact testing, as specified above: for the particular event, no alternative event can take place simultaneously.\\
Define $Pr(a_i)= \Psi(a_i)\Psi^{*}(a_i)\geq 0$.
In this case for $a_{1}\neq a_{2}$ 
\begin{eqnarray}
Pr(a_{1} \vee a_{2}) = \Psi(a_{1} \vee a_{2})\Psi^{*}(a_{1}\vee a_{2})=
\nonumber\\
<a_{1}|\Psi><\Psi|a_{1}>+<a_{2}|\Psi><\Psi|a_{2}>= \nonumber\\
Pr(a_{1}) + Pr(a_{2}). 
\label{prob_or}
\end{eqnarray}
The countinuous set of sates may be described in the similar fashion.

This together with (\ref{scal_1}) constitute the set of axioms, as defined for classical probability, so that the law of large numbers can be applied. Algebra of nonnegative numerical values corresponds to logic of events according to the set of rules used in the classical probability theory, the condition of exact testing is satisfied, that means that these numerical values can be identified with the appropriate probabilities of events, as conventionally defined.
The formally introduced value $Pr$ as the measure defined on a Banah space (an abstract mathematical value) does describe the characteristic of the physical process. It may be noted that the condition of exact testing, that is the interaction due to which only a single value of the particular parameter is prescribed to the system, has been postulated in [$\ddagger$]. The mutually excluding values of the parameter are defined in the process of interaction of the given type. This means the act of interaction conducted on the DI system would transform its state to the one, in which the standard probabilistic model applies. 
\\

In the general case of parameter $x$ we specify (compare with Youssef \cite{youssef})
\\

{\em Definition:} We call the function $\Psi(x)$ with the meaning and properties described above ``generalized probability" or in brief 
\\
``g-probability" of entering the state $x$.\footnote{We also will be using the term g-probability in relation to components of the function $U(x)$, $V(x)$.} 
\\

It may be noted that the classical system with continuous interactions is fundamentally described by a single set of states ${\bf A}$(different, mutually exclusive states correspond to different values of the parameter $A$), so the classical probability measure can be defined on the set ${\bf A}$.
The DI-system is described by the set ${\bf A}\oplus{\bf B}$: entering of the given state $a$ in between interactions is considered in relation to particular state of interaction $b_0$ (or the state $b$, which may belong to logically possible sequence of states between the given state and the state of interaction $<a|b><b|b_{0}>$).

Without an explicit reference to the process of interaction, the relation between g-probability and classical probability can be formally established, provided the mapping ${\bf A}\oplus{\bf B}\rightarrow {\bf A}$. 
In particular the relation between the g-probability and the classical probability ay be formally defined for the set of elementary states ${\bf A}$, according to the formula (\ref{prob}), if explicitly postulated the relation similar to (\ref{prob_or})
\begin{equation}
Pr(a_{i}\vee a_{j})= Pr(a_{i})+Pr(a_{j}).
\end{equation} 
Mathematically that means that we only define mapping for each single pair $<a_{i}|b>\in{\bf A}\oplus {\bf B}$ to $a_{i}\in {\bf A}$, not for $\sigma$-algebra specified for ${\bf A}\oplus {\bf B}$ (see Appendix).   
\\

The definition of DI-interaction, which in particular implied [$\ddagger$], was based on the assumption that elementary states exist and can be entered. That is, the states, which correspond to single values of the parameter, may be entered as a result of the process such that if a particular state is entered, no other state is entered. Now we specify the conditions necessary for existence of the states which allow such kind of process.  

\section{\rm Elementary states}

Consider two related representations of an intermediate state.
As defined above, the state specified as an elementary state, which corresponds to a given value $p$ of the parameter $P$, in terms of the other parameter $X$, can be presented by a set of elementary states described by the function $\Psi_{p}(x)$. This implies that the set of elementary states exists for the parameter $X$: we assume existence of the particular representation of an intermediate state and then discuss mathematical description of related representations. In the case of motion of the DI-particle in space, we assume that the set of positions constitute the particular representation of an intermediate state and then consider the related representations.

For the set of values $p_{i}$, the set of functions $\Psi_{p_{i}}\equiv \Psi_{i}$ is defined.
According to the definition, unconditional transitions between different elementary states are forbidden. In terms of functions $\Psi_{i}(x)$, that means that the functions with the different values of $i$ are orthogonal with respect of inner product
\begin{equation}
(\Psi_{i}\cdotp\Psi_{k}) \equiv \int_{X}\Psi_{i}(x)\Psi_{k}^{*}(x)dx.
\label{dot_product}
\end{equation}
Indeed the simple expression of orthogonality of the functions $\Psi_{p_{i}}$
\begin{equation}
\int_{X}<\Psi_{i}|x><x|\Psi_{k}>dx\equiv e_{i}\delta_{i,k},
\label{proc_1}
\end{equation}
($e_{i} = const$) interpreted as a definition of the process $\Psi_{i}\rightarrow\Psi_{j}$, implies that if the system enters the state $\Psi_{i}$, its state cannot also include  $\Psi_{j}$ and the system cannot pass into $\Psi_{j}$, for $i\neq j$. 

If the set of functions $\Psi_{i}(x)$ constitutes a complete set of orthogonal functions, any state defined for the given representation can be presented as a set of elementary states
$\Psi(x)=\sum c_{i}\Psi_{i},$
with the coefficients 
$c_{i} = \int_{X}\Psi(x)\Psi^{*}_{i}(x)dx = <\Psi|\Psi_{i}>,$
which have a meaning of g-probability of unconditional transition from $\Psi$ to $\Psi_{i}$.

In general, the set of orthogonal functions $\Psi_{i}$ may be obtained from any set of linearly independent functions using Gramm-Schmidt procedure. Alternatively, this set may be defined as linearly independent solutions of Sturm-Liouville problem; for the given type of boundary conditions this set of orthogonal functions is complete.

The set of orthogonal functions ${\Psi_{k}}$ allows us to construct a Hermitian 
operator
\begin{equation}
P(x,\eta)=\sum_{k}p(k)\Psi_{k}(x)\Psi_{k}^{*}(\eta)
\label{rr_1}
\end{equation}
for discrete $k$ and
\begin{equation}
P(x,\eta)=\int_{k}p(k)\Psi_{k}(x)\Psi_{k}^{*}(\eta)dk
\label{rr_2}
\end{equation}
for continuous $k$; $p(k)$ is a real function.
The functions $\Psi_{k}(x)$ are eigenfunctions of the operator
$\hat{P}$, $\hat{P}\Psi(x)\equiv\int P(x, \eta)\Psi(\eta)d\eta$, which correspond to the eigenvalues $p(k)$. 

For operator defined according to (\ref{rr_1}), (\ref{rr_2}) and an arbitrary g-probability \\
 $\Psi(x)=\sum_{j}c_{j}\Psi_{j}(x)$ we have
\begin{equation}
\int_{x}\Psi^{*}(x)\hat{P}\Psi(x)=\bar{p},
\label{p_bar}
\end{equation}
with $\bar{p}=\sum_{k} p_{k}<\Psi_{k}|\Psi><\Psi|\Psi_{k}>$ defined as an average value in probabilistic sense, assumed the relation of g-probability and classical probability as defined above. 

Generally, the set of square integrable complex functions of the real parameter, with inner product specified according to (\ref{dot_product}) will constitute a Hilbert space. We can specify a Hermitian operator defined by the basis vectors of the Hilbert space according to (\ref{rr_1}), (\ref{rr_2}).
For formally introduced set of orthogonal functions $\Psi_{k}$, which will define a Hermitian operator, we specify the function $a(k)$ which may be mapped to the particular parameter. In this case we can relate the operator to this parameter; elementary states of the DI-system $\Psi_{k}$ are defined by the values of the parameter $a=a(k)$.
 
Alternatively the set of orthogonal functions specified by the particular Hermitian operator, is assumed to relate to elementary states, as defined for the given parameter. Recalling that for Hermitian operators, eigenfunctions which correspond to different eigenvalues constitute complete set of orthogonal functions, the set of functions which represent elementary states of the parameter, may be defined as the set of eigenfunctions for the given Hermitian operator. 

Note that to define the functional Hilbert space which specifies the given representation of an intermediate state $P$, we assumed existence of the initial set of elementary states $X$. This case also can be described within the terms of the current formalism, with elementary states $x_{k}\in E^{1}$ represented by delta function $\delta(x-x_{k})$, consequently, the operator which represents the parameter will be equal to $x$.

Generally, if postulate motion of the DI-system in the vector space $A$, so that 
\\
g-probability of entering of an elementary state $a_{k}\in A$ is defined by $\delta(a-a_{k})$, the complete set of orthogonal, square integrable complex functions $\Psi_{b}(a)$ will specify the related representation of an intermediate state, which corresponds to the parameter $b$, with the average value of $b$ defined as
\[
\bar{b}= \int_{A}\Psi^{*}(a)\hat{B}\Psi(a)da,
\]
where
\[
\hat{B}=\int_{b}b\Psi_{b}^{*}(a)\Psi_{b}(a_{1})db.
\]

\section{\rm The physical parameters and representations of the intermediate
state}
When formulated the basic statements, we defined that elementary states in the given representation should correspond to the particular values of a given classical physical parameter.
On the other hand, we defined an eigenvalue of the particular Hermitian operator $\hat{P}$ as a kind of label (or a function of this label), which identifies the particular elementary state $\Psi_{p}$.
According to the definition, detecting of the system in the appropriate elementary state is described by the particular value $p$ which is an eigenvalue of the operator $\hat{P}$. 
The definitions prove to be consistent if we are able to show, how the particular values of the parameter, defined for the classical system, can be mapped to the eigenvalues of the operator $\hat{P}$ (the states of the DI-system when the states are detected).
That is, $\bar{p}$ defined by (\ref{p_bar}) is an average value of the classical physical parameter $P$ attributed to the state of DI-system $\Psi(x)$.

For integrals of motion defined for the classical particle, we consider whether the appropriate parameters will specify the sets of elementary states for the DI-particle. 

\small
\subsection{\rm One dimensional motion}
We continue with example of motion of the particle and consider the operator, which describes transition of the particle between the different positions in one-dimensional case. Transition between the positions (defined by the values of coordinates) means, that from the state which includes a given point $x$, the particle should pass to the new state, which does not include $x$.

The g-probability to pass into the state which does not include the given position $x$ will be given by the expression 
\begin{equation}
\Pi=\int_{-\infty}^{\infty}\Psi(x_{1})\phi_{1}(x,x_{1})dx_{1},
\label{27}
\end{equation}
$\Psi(x_{1})$ is g-probability that the system enters the state, which includes $x_{1}$ in vicinity of $x$, $\phi_{1}(x_{1},x)$ is g-probability of transition from $x_{1}$ to $x$ (${\cal V}_{x_{1},x}$ process, ordering between the states is defined), or more generally that the state which includes $x$ does not include $x_{1}$. 

According to the definition $\phi_{1}(x_{1},x)=-\phi_{1}(x,x_{1})$, or for
$\phi(x,\epsilon)=\phi_{1}(x,x_{1})$,
($\epsilon = x - x_{1}$), in the case of homogeneous space
\begin{equation}
\phi(\epsilon)=-\phi(-\epsilon).
\label{28a}
\end{equation}

Assume that for the particle, if {\em the state includes the position $x$, g-probability that the particle passes to the position $x_{1}=x+\epsilon$ rapidly decreases with the growth of $\epsilon$.} 
Recalling (\ref{28a}), after substitution of the expansion
$\Psi(x+\epsilon)
=\Psi(x)+\epsilon\frac{d\Psi}{dx}+
\frac{\epsilon^{2}}{2}\frac{d^{2}\Psi}{dx^{2}}+o(\epsilon^{2})$ we obtain 
\begin{equation}
\Pi=\frac{d\Psi}{dx}\int_{-\infty}^{\infty}\phi(\epsilon)\epsilon d\epsilon
+ \int_{-\infty}^{\infty}o(\epsilon^{2})d\epsilon.
\end{equation}
Assumed the integrals exist and the second integral is small compared with the first, we have an asymptotic expression
\begin{equation}
\Pi\simeq\frac{d\Psi}{dx}\int_{-\infty}^{\infty}\phi(\epsilon)\epsilon d\epsilon=i\frac{\bar{x}}{\hbar}\hat{p}\Psi,
\end{equation}
here $\bar{x}=\Bigl|\int_{-\infty}^{\infty}\phi(\epsilon)\epsilon d\epsilon\Bigr|$, and
we defined the operator
\begin{equation}
\hat{p}=-i\hbar\frac{d}{dx},
\label{30}
\end{equation}
which specifies transitions between the states $x$. The factor $-i$ is used to define a Hermitian operator (this implies that $Re \phi\equiv 0$, which is fully consistent with the definition of ${\cal V}$-process).
According to its definition the operator $\hat{p}$  transforms the g-probability in the initial coordinate representation, to the function which characterizes the transition between different positions.
To make notations similar to those, used in the conventional quantum theory we also added a small constant multiple $\hbar$. 

Eigenfunctions of the Hermitian operator\footnote{the right hand side denotes an operator; $\Psi(x)\rightarrow 0$, when $x\rightarrow \pm \infty$} 
\\
$-i\hbar\frac{d}{dx}\equiv \int d\eta\int_{-\infty}^{\infty}pe^{ip(x-\eta)/\hbar}dp,$
defined for $x\in(-\infty, \infty)$ 
\begin{equation}
\Psi_{p}(x)=e^{ipx/\hbar},
\end{equation}
constitute a complete set of orthogonal functions; the functions $\Psi_{p}(x)$ describe elementary states of a measurable parameter $p$, with a meaning of a number of transitions between the positions per unit length.

\large
\subsection{\rm Parameters of motion of the particle}

The similar approach may be applied to an arbitrary parameter of the system.  
Consider the set of states of the system described by a vector space. Define automorphism of this vector space, as specified by a given Lie group ${\cal L}$. Consider the particular trajectory $\Gamma$, in the vector space, which is invariant under transformation defined by one parametric subgroup of the group ${\cal L}_{t}$ and parameterized by the variable $t$. 

G-probability to pass between the states on the trajectory, which means to enter the state which includes $t_{1}$, provided the system passes from $t_{1}$ to $t$, or more generally, the state which includes $t_{1}$ does not include $t$, is defined by the formula
\begin{equation}
\Pi = \int_{\Gamma}\phi_{1}(t, t_{1})\Psi(t_{1})dt_{1}
\end{equation}
According to the definition, $\phi_{1}$ is an
antisymmetric function of its arguments
$\phi_{1}(t,t_{1})=-\phi_{1}(t_{1},t)$.
For
$\phi(t,\epsilon)=\phi_{1}(t,t+\epsilon)$,
due to the symmetry of the vector space, as defined by the Lie group, the function $\phi$ does not explicitly depend on $t$. In the case if $\phi(\epsilon)$ is a quickly decreasing function  
\begin{eqnarray}
\phi(\epsilon) \epsilon^{n}\rightarrow 0, & {\rm for} & \epsilon\rightarrow 0,
\end{eqnarray}
we can use an expansion of $\Psi(t)$ along the trajectory $\Gamma$.
\begin{equation}
\Psi(t_{1}) = \Psi(t) + L_{t}\Psi(t)\epsilon + o(\epsilon),
\end{equation}
here $L_{t}\Psi$ is Lie derivative which corresponds to the given subgroup 
of ${\cal L}$.
Then we have
\begin{eqnarray}
\Pi\simeq L_{t}\Psi\int_{\Gamma}\phi(\epsilon)\epsilon d\epsilon=cL_{t}\Psi,
\;\; c=const.
\end{eqnarray}
If the expression can be transformed to a Hermitian operator $\hat{L}\equiv cL_{t}=\int k\Psi_{k}\Psi^{*}_{k}dk$ by the appropriate choice of the constant $c$, we can define a measurable physical parameter $l$, which describes the given type of transitions between the states, so that 
elementary states are defined as eigenfunctions $\Psi_{l}$ of the operator $\hat{L}$ and the values of the parameter correspond to the eigenvalues $l$ of the operator. 
\\

Consider examples of transformations of $E^{3}$, which describe transition of the particle between different positions in space, as defined by the trajectory $\Gamma$. That is, the trajectory is invariant under transformations of the group $L_{t}$; $t$ is the parameter of transformation.
We define the process of transition of the particle to the state ${\bf r}$ as entering the state ${\bf r}_{1}$ in vicinity of ${\bf r}$, provided the system passes from ${\bf r}_{1}$ to ${\bf r}$; so that ${\bf r}(t)\in \Gamma$, 
${\bf r}={\bf r}(t_{0})$, ${\bf r}_{1}={\bf r}(t)$.
The g-probability of the process 
\begin{equation}
\Pi({\bf r}) = \int_{\Gamma}\phi_{\Gamma}({\bf r},{\bf r}_{1}(t))
\Psi({\bf r_{1}}(t))dt,
\label{rrr}
\end{equation}
here $\Psi({\bf r_{1}})$ is g-probability to enter the position ${\bf r_{1}}$,
$\phi_{\Gamma}$ defines g-probability that the state which includes
${\bf r_{1}}$ does not include ${\bf r}$, an antisymmetric function of its arguments. For the case of homogeneous, isotropic space $\phi_{\Gamma}({\bf r},t)= \phi(t)$, so that $\phi(t)=-\phi(-t)$; in this case 
\begin{equation}
\Pi({\bf r}) \simeq \Bigl[\int_{\Gamma}\phi(t)dt\Bigr]\frac{dr_{i}}{dt}\frac{\partial}{\partial r_{i}}\Psi({\bf r})
\label{rrr1}
\end{equation}
\\

{\em Momentum.}

Consider motion in $E^{3}$ along the given direction ${\bf n}$ ($\|{\bf n}\|=1$).
Lie derivative which specifies transformation of the scalar function $\Psi$ is defined by ${\bf n}\nabla$; the Hermitian operator which describes motion in the given direction can be defined using the same expression multiplied by $i$.

The similar result can be obtained directly if consider g-probability
of transition in space along the direction ${\bf n}$. 
Substitution of
${\bf r}={\bf r}_{0}+{\bf n}t$, 
into (\ref{rrr1}) gives
\[
\Pi\simeq{\bf n}\nabla\Psi \int_{-\infty}^{\infty}\phi(t)tdt.
\]
Taking into account that ${\bf n}$ is an arbitrary unit vector, we define a Hermitian operator 
\begin{equation}
\hat{p}=-i\hbar\nabla,
\end{equation}
which should relate to a measurable parameter ${\bf p}$. The equation
$(-i\hbar\nabla -{\bf p})\Psi({\bf r}) = 0$
specifies a measurable parameter, which describes transitions between different positions in space.
According to its definition, 
\\
${\bf p}=(p_{x}, p_{y}, p_{z})$ is a vector, components of which specify the average number of transitions of a particle between the states with different positions at a unit length (along the orthogonal directions in space).

Recalling the condition that the expression for the commutator of Hermitian operators \\
$[\hat{A},\hat{B}]=i\hat{C}$ implies the uncertainty relation 
\\
$(\Delta A)(\Delta B) \geq \frac{1}{2}|<\hat{C}>|$, we can establish Heisenberg's uncertainty principle for coordinates and components of momentum
\begin{equation}
\Delta x \Delta p_{x} \geq \frac{\hbar}{2} \ldots,
\end{equation}
which means that the exact values of coordinate and momentum cannot be specified simultaneously, in full agreement with our definition of momentum as characteristic of transition between the positions.
\\

{\em Rotational momentum.}
\\

Consider rotation in isotropic space, which is described by $SO(3)$ transformation of $E^{3}$. 
The Lie derivative which corresponds to rotation is defined as ${\bf n}[{\bf r}\nabla]$, so that the Hermitian operator which defines transitions between the states, may be presented as
$i[{\bf r}\nabla]$.

The result can obtained by substitution of 
${\bf r}(t)={\bf r}_{0}+[{\bf r}{\bf n}]t$, $t\in [-\pi,\pi]$, into (\ref{rrr1}), which gives  
\begin{equation}
\Pi \simeq {\bf n}
[{\bf r}\nabla]\Psi({\bf r})
\int_{-\pi}^{\pi}\phi(t)t dt,
\end{equation}
assumed that
$\phi(t)=\phi(t + 2\pi)$.
Recalling that ${\bf n}$ is a constant vector, we define the Hermitian operator
\begin{equation}
\hat{M}=-i\hbar[{\bf r}\nabla].
\label{230}
\end{equation}
which we call the operator of rotational momentum. 
The equation
$(-i\hbar[{\bf r}\nabla]-{\bf M})\Psi=0$
defines the measurable parameter ${\bf M}$ which is a vector with the discrete set of values for its components. Components of ${\bf M}=(M_{x}, M_{y}, M_{z})$ describe the transitions of the particle between the positions with the different angular coordinates. 
\\

Transitions between the internal states, described by the variables not related to space coordinates (supposed such states and variables exist) may be considered in the same way.
As demonstrated, for a variable, which describes a state of the system, we can define an operator which specifies transitions between the states, if consider the appropriate symmetry group of the system.
\\

{\em Internal U(1) transformation.}
\\

The Hermitian operator defined by Lie derivative which corresponds to $U(1)$ transformation of $C^{1}$ is  $i\frac{\partial}{\partial\nu}$. Consider this in more detail.
Assume that the state of the particle is described both by space coordinates and an internal parameter $\nu$, provided
\\
$\Psi({\bf r},\nu)=\psi_{r}({\bf r})\psi(\nu)$,
$\psi(\nu) = \psi(\nu + 2\pi/q)$,
$q=const$. Similarly to the case of transitions in space define that transition into the state $\nu$ means entering of the state $\nu_{1}$, provided the system passes from  $\nu_{1}$ to $\nu$.
The g-probability of the process, similarly to (\ref{27}), is
\begin{equation}
\Pi \simeq \int_{-\pi/q}^{\pi/q}\phi_{1}(\nu,\nu_{1})\psi(\nu_{1})d\nu_{1},
\label{nu}
\end{equation}
here $\psi({\nu})$ is g-probability to enter the state ${\nu}$,
$\phi_{1}$ defines g-probability that the state which includes
$\nu$ does not include $\nu_{1}$, an anti-symmetric function of its arguments
$\nu,\nu_1\in[-\pi/q, \pi/q]$.

Proceeding in the customary manner, we obtain the equation 
$\Bigl(-i \hbar\frac{\partial}{\partial\nu} - q_{n} \Bigl)\phi_{n} = 0,$
which specifies the eigenfunctions of the Hermitian operator;
recalling that $\phi(\nu)$ is a periodic function, we have
$q_{n}\in\{0, \pm q, \pm 2q, \dots \}$.
\\

The other internal symmetries can be considered in the similar fashion. 
However even without a detailed discussion of each case, we can make a conclusion, that if the domain of definition of the internal variable is finite, the set of values, which specifies the parameter of transitions, is discrete.

\section{\rm The behavior of DI-system in time}

As established above, for the particular elementary state
${\cal A}$ at the given moment in time, we can define a non-ordered state 
${\cal U}({\cal A})$ which is a set of elementary states entered simultaneously with ${\cal A}$ (ordering for ${\cal U}$ and ${\cal A}$ is not defined) and ordered state ${\cal V}({\cal A})$ -- a set of elementary states, such that the system passes from ${\cal V}$ to ${\cal A}$ or from ${\cal A}$ to ${\cal V}$ (ordering is defined for the states ${\cal A}$ and ${\cal V}$).
Now consider the behavior of the DI-system in time. To define the behavior of the system in time we assume that, similarly to the classical case
DI-system may change its state, and {\em the states of the system may be parameterized using an ordered variable $t$}.

In the present non relativistic model we use the conventional interpretation of time: time in the DI system is interpreted as an ordered variable, which is continuously prescribed to the system. 

Suppose that the system can appear in the state ${\cal V}({\cal A})$ at the moment of time $t+\delta t$ (pass from ${\cal A}$ to ${\cal V}$), if this system was staying in the state ${\cal U}({\cal A})$ at the moment $t$
(${\cal U}({\cal A})$ included ${\cal A}$). The system can appear in the state ${\cal U}({\cal A})$ at the moment of time $t+\delta t$, if it passed into this state from the state ${\cal V}({\cal A})$, defined for $t$. This specifies the relations between {\em ordering on the set of states} as defined for DI-system and {\em standard ordering in time}. 
The behavior of the system in time is defined as transitions between the sets of states ${\cal U}$, ${\cal V}$ (non-ordered and ordered states relative to ${\cal A}$), so that the system passes from ${\cal U}$ to ${\cal V}$ and from ${\cal V}$ to ${\cal U}$. 
In this case time has a meaning of the ordering variable, which identifies the subsequent states of the system ${\cal U}$ or ${\cal V}$.

We assume that the change in time $dt$ of g-probability to stay in 
${\cal A}$ is proportional to 
\\
g-probability of transition 
${\cal V}\rightarrow {\cal A}$ multiplied by $dt$:
{\em the change of g-probability to stay in the given elementary state at the particular moment of time is proportional to g-probability to pass to the non-ordered state which contains this elementary state}
\begin{equation}
dU(a) =c_{u} V(A)dt,
\label{uut}
\end{equation}
here g-probability of transition ${\cal V}\rightarrow {\cal A}$ is defined as g-probability to pass to any elementary state of the non-ordered state which includes ${\cal A}$.

The change in time $dt$ of g-probability to exit from ${\cal A}$ is proportional to g-probability initially to stay in ${\cal U}({\cal A})$ multiplied by $dt$:
{\em the change of g-probability to exit the given elementary state is proportional to g-probability initially to stay in the non-ordered state which contains this elementary state}
\begin{equation}
-dV(a) =c_{v} U(A)dt,
\label{vvt}
\end{equation}
$c_{u}, c_{v}$ are constant coefficients.
\\
Here g-probability to stay in ${\cal U}({\cal A})$ is defined as g-probability to stay at any elementary state of the non-ordered state which includes ${\cal A}$.\footnote{Minus in the left-hand side of (\ref{vvt}) indicates that the system, which initially stayed in ${\cal U}({\cal A})$, exits from ${\cal A}$.}
The coefficients $c_{u},c_{v}$ can be made equal to by the appropriate scaling of the functions $U$, $V$. Without loss of generality we set $c_{u}=c_{v}=1/\hbar$. 
The formulae (\ref{uut}), (\ref{vvt}) constitute the additional fundamental assumption, which should be added to the basic principles specified in the section 3.
\\
\\
{\em The system with discrete states.}

For the discrete set of elementary states, the change of g-probability to stay in the state $A_{i}$ is defined to be proportional to g-probability to pass into the non-ordered state which includes $A_{i}$. 
\begin{equation}
\hbar dU_{i}=H_{ij}V_{j}dt,
\label{51}
\end{equation}
similarly the change of g-probability to pass from the state $A_{i}$ is proportional to g-probability to stay in the non-ordered state which includes $A_{i}$
\begin{equation}
\hbar dV_{i}=-H_{ij}U_{j}dt,
\label{52}
\end{equation} 
here $U_{i}$ is the g-probability that the system stays in the state $A_{i}$, $V_{j}$ is the g-probability  for the system to pass into the state $A_{j}$, $H_{ij}$ is the probability that the intermediate state of the system which includes $A_{i}$ also includes $A_{j}$, or more generally, $H_{ij}$ is the g-probability that the intermediate state of the system which includes $A_{i}$ also includes $A_{j}$, or the system passes from the intermediate state which includes  $A_{i}$ to the intermediate state which includes $A_{j}$. 
The matrix $H_{ij}$ is Hermitian
$H_{ij}=H_{ji}^{*}$.
Combining (\ref{51}) and (\ref{52}) we obtain
\begin{equation}
i\hbar \frac{d\Psi_{i}}{dt}=H_{ij}\Psi_{j},
\end{equation}
which describes the behavior of the DI-system defined for the discrete set of states.

Now we establish the equation of motion for the particle in space and then similarly consider the behavior of the field.

\small
\subsection{\rm Motion of the particle in space} 
In one-dimensional case, elementary states of the particle are defined based on the values of its coordinate $x$.
Consider a particular elementary state $x$, which belongs to an intermediate state at the particular moment of time. 
We define $U(x)$ as the g-probability for a particle to stay at an elementary state $x$ and $V(x)$ the g-probability to pass into the state $x$. As stated above, the g-probability $U(x)$ changes in time due to possible transitions into this state; the change of $U(x)$ in unit time is  proportional to intensity of transitions. 

Similarly to the previous, we use the definition of the particle (in contrast to the field), as a physical entity with the property that if its state includes the position $x$, the g-probability $f(x)$, that this state also includes the position $x_{1}=x+\epsilon$, rapidly decreases with growth of $\epsilon$.

The g-probability to pass into the state which includes the position  $x$ can be presented as a g-probability $V(x_{1})$ to pass to the elementary state $x_{1}=x+\epsilon$, multiplied by the g-probability $f(x,x_{1})$ that the intermediate state, which includes $x_{1}$ also contains $x$.
According to the definition
$f(x,x_{1})=f(x_{1},x)$.
In further, instead of $f(x,x_{1})$ we use
$\phi(x, \epsilon)=f(x,x_{1})$,
$\epsilon=x-x_{1}$, so that
$\phi$ is an even function of $\epsilon$.
We assume that $\phi$ is a genralized function of the form
\begin{equation}
\phi(x, \epsilon) = r_{0}(x)\delta(\epsilon) + \hbar^{2} r_{2}(x,\epsilon) + o(\hbar^{2}),
\label{4}
\end{equation}
$\delta()$ is Kroneker $\delta$-function, and $r_{0},r_{2}$ are classical functions. Note that for the classical system $\phi(x, \epsilon)= 0$ if $\epsilon\neq 0$; the second term in the formula is specific for the DI-system.

For g-probability $U(x)$ we have
\begin{equation}
\hbar dU(x)=dt\int_{-\infty}^{\infty}V(x+\epsilon)\phi(x,\epsilon)d\epsilon.
\label{Schr0U}
\end{equation}
Recalling that for the particle, the function $\phi(x,\epsilon)$ is rapidly decreasing with the growth of $\epsilon$, we can expand $V(x+\epsilon)$ in the vicinity of $x$, substitution of the first two non vanishing terms gives
\begin{eqnarray}
\hbar dU \simeq \;\;\;\;\;\;\;\;&\;&
\nonumber\\ 
dt\Bigl[V\int_{-\infty}^{\infty}\phi(\epsilon)d\epsilon 
&+&
\frac{1}{2}\frac{\partial^{2}V}{\partial x^{2}}\int_{-\infty}^{\infty}
\epsilon^{2}\phi(\epsilon)d\epsilon  + 
o(\epsilon^{3})\Bigr].
\end{eqnarray}
In the notations
\[
P(x) = \int_{-\infty}^{\infty}\phi(x, \epsilon)d\epsilon
= r_{0}(x) + 2\hbar^{2}\int_{0}^{\infty}r_{2}(x,\epsilon)d\epsilon,
\]
\[
Q(x) = \frac{1}{2}\int_{-\infty}^{\infty}\phi(x, \epsilon)\epsilon^{2}d\epsilon
= \hbar^{2}\int_{0}^{\infty}r_{2}(x,\epsilon)\epsilon^{2}d\epsilon,
\]
we have
\begin{equation}
\hbar\dot{U}= P(x)V+Q(x)\frac{\partial^{2}V}{\partial x^{2}}
\end{equation}
For the case of the homogeneous space $\phi$ does not depend on $x$, so $P, Q$ are constants. More generally, consider the case when $r_{0}$ in (\ref{4}) depends on $x$, in the notations 
\begin{eqnarray}
Q = -\frac{\hbar^{2}}{2m},
\;\;
P(x) = R(x),
\label{m}
\end{eqnarray} 
we have
\begin{equation}
\hbar\dot{U}(x)=-\frac{\hbar^{2}}{2m}\frac{\partial^{2}V(x)}{\partial x^{2}}+
R(x)V(x),
\label{R}
\end{equation}
Now consider the change in time of g-probability of transitions between the states $V(x)$. To pass form the state $x$ at the moment $t=t_{0} + dt$ the particle should stay in the state $x$ at the moment $t=t_{0}$. The g-probability to stay in the state which includes the position $x$ can be presented as the g-probability to stay in $x_{1} = x+\epsilon$ multiplied by the g-probability $f(x,x_{1})$ that the state which includes $x_{1}$ also contains $x$.
Similarly to the previous case we use
$\phi(x,\delta) = f(x,x_{1})$.
Substitution into (\ref{vvt}) gives 
\begin{equation}
-\hbar dV=dt\int_{-\infty}^{\infty}U(x+\delta)\phi(x,\delta)d\delta,
\label{Schr0V}
\end{equation}
so that after expansion in the vicinity of $x$, using the notations
(\ref{m}), we have
\begin{equation}
-\hbar \dot{V}(x)=-\frac{\hbar^{2}}{2m}\frac{d^{2}U(x)}{dx^{2}}+
R(x)U(x).
\label{I}
\end{equation}
After re-scaling, the expressions (\ref{R}) and (\ref{I}) can be written as one formula for the complex function $\Psi(x)=U(x)+iV(x)$, 
\begin{equation}
i{\hbar}\dot{\Psi}_{t}=-\frac{{\hbar}^{2}}{2m}\Psi_{xx}+R\Psi.
\label{Schr}
\end{equation}
That is one dimensional Schr\"{o}dinger equation for the particle.
\\
\\
\large
The equation (\ref{Schr}) may be generalized and established for the case of the three dimensional motion of the particle in $E^{3}$, if assume that transitions between the coordinates in orthogonal directions take place independently.
We write the equations of the type (\ref{Schr0U}), (\ref{Schr0V}) already in the complex form for $\Psi\equiv \Psi({\bf r})$
\begin{equation}
i\hbar\dot{\Psi}({\bf r})=\int_{V}
f({\bf r}, {\bf r}_{1})\Psi({\bf r}_{1})d{\bf r}_{1},
\label{Schr30}
\end{equation}
$f({\bf r},{\bf r}_{1})$ is g-probability that if the state of the particle includes the position ${\bf r}$, it also contains the position ${\bf r}_{1}$.
According to its definition $f$ is symmetric function of its arguments
\\
$f({\bf r},{\bf r}_{1})=f({\bf r}_{1},{\bf r})$.
For $\phi({\bf r},\delta{\bf r})=f({\bf r},{\bf r}_{1})$,  
$\delta{\bf r}={\bf r}_{1}-{\bf r}$, 
we have
\begin{equation}
\phi({\bf r},\delta{\bf r})=
\phi({\bf r},-\delta{\bf r}).
\end{equation}
Recalling that the function $\phi(\delta{\bf r})$ is rapidly decreasing with growth of $\delta{\bf r}$, we rewrite (\ref{Schr30}) as
\begin{eqnarray}
i\dot{\Psi}({\bf r})=
\int_{V}
\phi({\bf r},\delta{\bf r})
\Bigl(\Psi({\bf r}) + \delta x_{k} 
\frac{\partial\Psi({\bf r})}{\partial x_{k}}+ \nonumber\\
\frac{1}{2}\delta x_{k}\delta x_{j}\frac{\partial^{2}\Psi({\bf r})}
{\partial x_{k}\partial x_{j}}
+o((\delta x_{k})^{3})
\Bigr)d^{3}(\delta x_{i}).
\label{Schr31}
\end{eqnarray}
Supposed that the space is locally isotropic, that means that in the vicinity of the particular point ${\bf r}$ the function $\phi$ depends only on the distance between two positions of the particle:
$\phi({\bf r},\delta{\bf r})=\phi({\bf r},\delta r)$,
After integration, in the notations similar to one-dimensional case, we obtain
\begin{equation}
i\hbar\dot{\Psi}=-\frac{{\hbar}^{2}}{2m}\Delta\Psi+R\Psi,
\label{Schr3}
\end{equation}
this is the three dimensional Schr\"{o}dinger equation for the particle. 

Define operator $\hat{H}$ as
\begin{equation}
\hat{H}=\Bigl(-\frac{\hat{p}^{2}}{2m}+R\Bigr).
\end{equation}
$\hat{H}$ is a Hermitian operator, so will correspond to a measurable value $E$.
According to (\ref{Schr3}), the g-probability of entering the state with the given value $E$ can be presented as
\[
\Psi({\bf r}, t)=e^{-iEt/\hbar}\psi({\bf r}).
\]

\subsection{\rm Classical system: Physical parameters}
\small
Similarly to quantum mechanics, for each particular operator $\hat{x}$, we define the operator $\hat{\dot{x}}=i[\hat{H}\hat{x}]$, so we have 
\begin{equation}
\hat{\dot{\bf p}}=-\nabla R(\hat{\bf r}),
\label{Newton1}
\end{equation}
which has the form of the classical Newton equation written for operators. 

Considering the mean values $<\hat{\bf p}>$, $<\hat{\dot{\bf p}}>$ and referring to Ehrenfest theorem, we can establish the classical system specified by the set of Hamilton equations. Alternatively, we can consider Feynman quantization.  
The equation (\ref{Schr3}) defined as equation of motion of the DI-particle is the Schr\"{o}dinger equation, so all mathematical apparatus developed in the quantum theory may be applied to the DI-system.
In particular partial solution of (\ref{Schr3}), 
$\Gamma(t_{1}, t_{2})\equiv e^{-\frac{i}{\hbar}\int_{t_{1}}^{t_{2}}\hat{H}dt}
\delta({\bf r}-{\bf r}(t_{1}))\delta(t-t_{1})$,
that is the standard quantum mechanical propagator for the particle in space 
\begin{equation}
\Gamma(t_{1},t_{2}) = 
\int D{\bf r}\exp\frac{i}{\hbar}\int_{t_{1}}^{t_{2}} L({\bf r},t)dt,
\label{pathint}
\end{equation}
here $L({\bf r},t)$ is defined as
\begin{equation}
L({\bf r},t) = \frac{m}{2}\dot{\bf r}^{2} - R({\bf r}),
\label{lagrangian}
\end{equation}

The formula may be used to define the classical system which corresponds to the given DI-system (in our case the DI-particle).
The definition (\ref{4}) implies 
$\phi(x,\epsilon)\rightarrow r_{0}\delta(\epsilon)$ if 
$\hbar\rightarrow 0$, that means the state of the system cannot contain  non equivalent elementary states. This is similar to the behavior of the classical system. 
In the asymptotic case when $\hbar\rightarrow 0$,
the integral (\ref{pathint}) can be presented as 
\[
\Gamma(t_{1},t_{2}) \sim 
\exp\frac{i}{\hbar}\int_{t_{1}}^{t_{2}} L_{0}({\bf r},t)dt, \;\; \hbar\rightarrow 0,
\]
where $L_{0}$ is defined according to the formula  
\begin{equation}
\delta \int_{t_{1}}^{t_{2}}  L_{0}({\bf r},t)dt = 0,
\label{leastaction}
\end{equation}
here $\delta$ denotes variation.
The formula (\ref{leastaction}) is the expression of the principle of the least action.
We may conclude that the classical system, which corresponds to the DI-particle, is described by the principle of the least action. This can be generalized for the case of an arbitrary DI-system, provided the condition similar to (\ref{4}) holds.

Based on the principle of the least action and properties of symmetry of space we can define the integrals of motion of the system, such as momentum
${\bf p}=\frac{\partial L_{0}}{\partial \dot{{\bf r}}}$, so that
the equation of motion for the classical particle obtained from
(\ref{lagrangian}), (\ref{leastaction}) has the form
\begin{equation}
{\dot{\bf p}}=-\nabla R,
\label{ClNewton}
\end{equation}
similar to (\ref{Newton1}). This may be interpreted as an indication that the operator $\hat{\dot{\bf p}}$ tends to the classical parameter 
$\dot{{\bf p}}$, when $\hbar$ 
tends to zero: $\hat{\bf p}\rightarrow {\bf p} + {\bf c}$,
if $\hbar\rightarrow 0$; the constant ${\bf c}$ may be chosen equal to zero.
Consider the equation 
\begin{equation}
(\hat{\bf p}-{\bf p}_{0})\Psi_{{\bf p}_{0}}(x) = 0,
\end{equation}
which specifies an elementary state of the particle defined by the eigenvalue ${\bf p}_{0}$ of the operator $\hat{\bf p}$.
In limit $\hbar\rightarrow 0$, the expression transforms to
\begin{equation}
({\bf p}-{\bf p}_{0})\Psi_{{\bf p}_{0}}(x) = 0;
\end{equation}
for the state $\Psi_{{\bf p}_{0}}(x)$, the value of ${\bf p}$ is equal to ${\bf p}_{0}$.
That means that the detection of the particle in the particular elementary state has the meaning of measuring of the parameter ${\bf p}$, which in this respect is related to the given operator $\hat{\bf p}$.
The relation is defined for operator of momentum, but may be extended to all operators presented as functions of momentum and coordinates, such as rotational momentum, or energy.\footnote{Note that within the framework of the conventional quantum theory, we cannot define the equation of motion for the classical particle, but only postulate that the form of the quantum equation for operators is similar to the corresponding classical equation.}

In our previous discussion we defined the states of the DI-system as related to the parameters of the classical system.
Now we are able to relate the states of the DI-particle not to the parameters of {\em ad hoc} chosen classical system, but infer them directly from the equation of motion of the DI-particle itself. Provided (\ref{4}), the integrals of motion obtained from the DI-equation in limit 
$\hbar\rightarrow 0$ specify the set of variables sufficient to define the parameters and then the states of the DI-particle itself. 

Note that it is impossible to give a description of any system without use of initial set of variables, existence of which is postulated, such as coordinates and time. However the other parameters (momentum, energy, rotational momentum) may be obtained using equation of motion of the DI-system without special reference to the corresponding classical system.
\\

\large
To establish equation of motion of an arbitrary DI-system (non necessarily the particle) the general approach described above can be used. Consider the set of states of the system parameterized by the variable $A$ and assume that for any elementary state $a$, the value of g-probability to stay in this state is changed in time due to transition to this state from the other states according to the formula
\begin{equation}
i\dot{\Psi}(a)=\int_{A}\theta(a,a_{1})\Psi(a_{1})d a_{1},
\label{schr_basic}
\end{equation}
here $\Psi(a)$ is g-probability of entering of the state $a$ and $\theta(a,a_{1})$ is a g-probability that the intermediate state which includes $a_{1}$ also contains $a$. 

\large
\subsection{\rm Schr\"{o}dinger equation for Grassmann variable}
We established equation of motion of the particle defined at the set of states, which are parameterized by the variable ${\bf r}=\{r_{j}\}$, where $r_{j}$ are $c$-numbers. In the similar fashion we can define equation of motion for the set of states parameterized by the set of Grassmann variables ${\bf b}=\{b_{j}\}$.

Assuming that the set of components of the Grassmann vector ${\bf b}$ represents the particular elementary state, behavior of the system may be presented as transition between the states in the vector space ${\bf B}$ similar to motion of the particle in space.
In this case for each particular moment in time we define g-probability
of entering or transition between elementary states  in the vector space  ${\bf B}$ as $\Psi({\bf b})\equiv U({\bf b}) + iV({\bf b})$ with the algebra defined by the formulae (\ref{gpv1}-\ref{gpv}).
The behavior of the system in time is defined by the equation of the type (\ref{schr_basic})
\begin{equation}
i\dot{\Psi}({\bf b})=\int_{\bf B}
\theta({\bf b}, {\bf b}_{1})\Psi({\bf b}_{1})d{\bf b}_{1},
\label{gras_Schr}
\end{equation}
here $\theta({\bf b}, {\bf b}_{1})$  is g-probability that if the state of the particle includes an elementary state ${\bf b}$, it also contains an elementary state ${\bf b}_{1}$. According to the definition 
\[
\theta({\bf b}, {\bf b}_{1}) = \theta({\bf b}_{1}, {\bf b}).
\]
Consider the particular case of two-dimensional Grassmann space ${\bf b}=\{b,b^{*}\}$. In this case we have
\begin{eqnarray}
\theta({\bf b},{\bf b}_{1}) &=& c_{00}+c_{11}(b+b_{1})+
c_{12}(b^{*}+b^{*}_{1}) \nonumber\\
&+&
c_{20}(b+b_{1})(b^{*}+b_{1}^{*}) \nonumber\\
&+& 
c_{30}(bb^{*}(b_{1}+b_{1}^{*})\nonumber\\
&+& 
b_{1}b_{1}^{*}(b+b^{*}))+
c_{40}bb^{*}b_{1}b_{1}^{*}, \nonumber
\end{eqnarray}
here $c_{ij}$ are constants.
Substitution into the integral (\ref{gras_Schr}) gives 
\begin{equation}
i\dot{\Psi}(b)=\hat{H}({\bf b})\Psi({\bf b}),
\label{Sch_Grassmannn}
\end{equation}
with operator $\hat{H}$ defined as 
\begin{eqnarray}
\hat{H} &=&
c_{00}\frac{\partial}{\partial b^{*}}
\frac{\partial}{\partial b}+
c_{11}\frac{\partial}{\partial b^{*}}+
c_{12}\frac{\partial}{\partial b} \nonumber\\
&+& 
c_{20} +
c_{30}(b+b^{*})+ 
c_{40}bb^{*}.
\label{ham_Grassmann}
\end{eqnarray}

Similarly to the case of motion of a particle in space, we define the operators $\hat{\bf b}$, 
$\hat{\dot{\bf b}}$.
For the case $c_{1j}=c_{30}=0$ we have (the standard formula of quantum theory applies)
\begin{equation}
\hat{\dot{\bf b}}=i[\hat{H}\hat{\bf b}].
\label{gras_dot}
\end{equation}

Consider the case when 
$c_{00}=1$, $c_{40}=\omega^{2}$, all other coefficients are equal to zero. 
In this case the Schr\"{o}dinger equation (\ref{Sch_Grassmannn})
will have the form
\begin{equation}
i\dot{\Psi}(b,b^{*})=\Bigl[\frac{\partial}{\partial b^{*}}
\frac{\partial}{\partial b}+
\omega^{2}bb^{*}\Bigr]\Psi(b,b^{*});
\label{gras_schr}
\end{equation}
for operators $\hat{\dot{b}}$, $\hat{\dot{b}}^{*}$, recalling (\ref{gras_dot}) we obtain
\begin{eqnarray}
\hat{\dot{b}}=i\frac{\partial}{\partial b^{*}},\;\;\;
\hat{\dot{b}}^{*}=-i\frac{\partial}{\partial b}.
\label{gras_eq_mot}
\end{eqnarray}
Consequently for $\hat{\ddot{b}}$,  $\hat{\ddot{b}}^{*}$ we have
\begin{eqnarray}
\hat{\ddot{b}}=-\omega^{2}\hat{b},\;\;\;
\hat{\ddot{b}}^{*}=-\omega^{2}\hat{b}^{*}.
\end{eqnarray}
This may describe a physically interesting case of oscillator.

\section{\rm Linear field}

Consider the linear field $\phi({\bf x})$ and define equation of behavior (motion) of the DI-field in space and in time. The field by definition is an entity distributed in space even at the moment of interaction. Consider the state of the field, which is a distribution in space, solution of the field equation with the given symmetry, for the particular constant time. 
In the case of a field specified by a linear equation with the given boundary conditions, a parameterized set of orthogonal functions defines solutions of this equation. The parameter of these solutions, such as amplitude, can be used to define elementary states of the DI-field.
We suppose that the elementary state of the DI-field can be described by an infinite dimensional vector, which is a set of amplitudes of the different harmonics of the function $\phi$ or the linear combination $c_{0}\phi+c_{1}\dot{\phi}$, $c_{0}, c_{1}=const$.
\\

We define the equation of motion for the linear (free) DI-field for the typical cases of the fields with integer and semi-integer spins. 

\subsection{\rm Integer spin}
{\em Klein-Gordon field.}
\\

Consider real Klein-Gordon field $\phi({\bf x},t)$, as defined by the formula
\begin{equation}
\Box\phi + m^{2}\phi=0,
\label{KG}
\end{equation}
here $m$ is a nonnegative constant\footnote{unless specified, we use $c=1$, $\hbar=1$}.
Consider vector 
${\bf a}\equiv\{a_{\bf k}\}$ 
which specifies the set of amplitudes of the particular harmonics of the field, so that
\begin{eqnarray}
\phi({\bf x},t) &=& \int \frac{d{\bf k}}{(2\pi)^{3}}a_{\bf k}(t) e^{-i{\bf kx}}
\nonumber\\
&\equiv&
\frac{1}{2}\int \frac{d{\bf k}}{(2\pi)^{3}}\Bigl(a_{\bf k}(t) e^{-i{\bf kx}}+
a_{-\bf k}(t) e^{i{\bf kx}}\Bigr).\nonumber\\
\;
\label{kg_main_field}
\end{eqnarray}
According to the definition
\begin{eqnarray}
\ddot{a}_{\bf k} + \omega^{2}_{\bf k}a_{\bf k} = 0,
\label{A}
\end{eqnarray}
where
\begin{equation}
\omega^{2}_{\bf k}={\bf k}^{2}+m^{2}.
\label{omega}
\end{equation}
We consider $b_{\bf k}=(a_{\bf k}+a_{-{\bf k}})/2$, 
$c_{\bf k}=(a_{\bf k}-a_{-{\bf k}})/2$, so that
\begin{eqnarray}
\ddot{b}_{\bf k} + \omega^{2}_{\bf k}b_{\bf k} = 0,\;\;
\ddot{c}_{\bf k} + \omega^{2}_{\bf k}c_{\bf k} = 0.
\label{BC}
\end{eqnarray}
We assume that the set of components of 2n-dimensional vector, $n\rightarrow \infty$ $\{(c_{\bf k}, b_{\bf k})\}$, ${\bf k} \equiv \{k_{i}\}$, $i=1,2,3$, represents the particular elementary state of the DI-field, so that the behavior of the system may be presented as motion in the vector space 
$({\bf B}_{n}\oplus{\bf C}_{n})$. That means for each particular moment in time we can define g-probabilities $U({\bf b},
{\bf c})$ and $V({\bf b},{\bf c})$ of entering or transition between elementary states  in $({\bf B}_{n}\oplus{\bf C}_{n})$, with the algebra defined by the formulae (\ref{gpv1}-\ref{gpv}). Alternatively we can define the complex g-probability function $\Psi({\bf b},{\bf c})$ so that 
$U( {\bf b},{\bf c} )=Re\Psi({\bf b},{\bf c})$, 
$V({\bf b},{\bf c})=Im\Psi({\bf b},{\bf c})$.
We need to specify the dependence of the 
\\
g-probability $\Psi({\bf b},{\bf c})$ upon time.
According to our general approach
\[
i\dot{\Psi}( {\bf b},{\bf c})=
\int_{\bf A}\theta({\bf b},{\bf c}, {\bf b}_{1},{\bf c}_{1})
\Psi({\bf b}_{1},{\bf c}_{1})d{\bf b}_{1}d{\bf c}_{1}.
\]
Similarly to the case of the particle we assume that if an intermediate state of the system includes an elementary state defined by $({\bf b},{\bf c})$, the g-probability $\theta({\bf b},{\bf c}; {\bf b}_{1},{\bf c}_{1})$ that this intermediate state also includes the state defined by 
$({\bf b}_{1},{\bf c}_{1} )=({\bf b}+\delta {\bf b}, {\bf c}+\delta {\bf c})$ rapidly decreases with growth of $\delta{\bf b}$, $\delta{\bf c}$, which would imply the relation similar to (\ref{4}).
Consider the case 
\[
\theta({\bf b},{\bf c}, {\bf b}_{1},{\bf c}_{1})=\sum_{\bf k}
\Bigl(\theta_{\bf k}(b_{\bf k},b_{1,{\bf k}})+
\theta_{1,\bf k}(c_{\bf k},c_{1,{\bf k}})\Bigr).
\] 
According to the definition 
\\
$\theta_{\bf k}(b_{\bf k},b_{1,{\bf k}})=
\theta_{\bf k}(b_{1,{\bf k}},b_{\bf k})$,
\\
$\theta_{1,\bf k}(c_{\bf k},c_{1,{\bf k}})=
\theta_{1,\bf k}(c_{1,{\bf k}},c_{\bf k})$; 
we use
$\theta_{\bf k}(b_{\bf k},b_{1,{\bf k}})=
\theta_{\bf k}(\delta b_{\bf k})$,
$\theta_{1,\bf k}(c_{\bf k},c_{1,{\bf k}})=
\theta_{1,\bf k}(\delta c_{\bf k})$,
with the following scaling for $\theta_{\bf k}$, $\theta_{1,\bf k}$ 
\begin{eqnarray}
\int \theta_{\bf k}(\delta b_{\bf k})(\delta b_{\bf k})^{2} 
d(\delta b_{\bf k}) &=& 1, \;\;\;\nonumber\\
\int \theta_{1,\bf k}(\delta c_{\bf k})(\delta c_{\bf k})^{2} 
d(\delta c_{\bf k}) &=& -1.
\end{eqnarray}
Assume that the topology of the vector space $({\bf B}_{n}\oplus{\bf C}_{n})$ is $E^{2n}$, $n\rightarrow \infty$, that means that we can repeat all arguments previously used to describe behavior of the particle and establish the equation of motion of the system in the form 
\begin{equation}
i\dot{\Psi}=  \sum_{\bf k}\Bigl[
-\frac{1}{2}\frac{\partial^{2}}{\partial b^{2}_{\bf k}} 
+\frac{1}{2} 
\frac{\partial^{2}}{\partial c^{2}_{\bf k}}
+R( b_{\bf k}, c_{\bf k} ) \Bigr]\Psi.
\label{SchrBC}
\end{equation}
To obtain the equation of motion for the field we need to provide that the constraint (\ref{BC}) holds in the classical limit, that is in the case if 
\\
$R(b_{\bf k},c_{\bf k}) =(1/2)\sum_{\bf k} \omega^{2}_{\bf k}
(b^{2}_{\bf k} - c^{2}_{\bf k})$, so we have
\begin{equation}
i\dot{\Psi}=\frac{1}{2} \sum_{\bf k} \Bigl[
-\frac{\partial^{2}}{\partial b^{2}_{\bf k}}+
\frac{\partial^{2}}{\partial c^{2}_{\bf k}}+
\omega^{2}_{\bf k} (b^{2}_{\bf k}-c^{2}_{\bf k})\Bigr]\Psi,
\label{SchrBC1}
\end{equation}
alternatively, using initial set of variables $a_{\bf k}$, we have
\begin{equation}
i\dot{\Psi}=\frac{1}{2} \sum_{\bf k} \Bigl[
-\frac{\partial}{\partial a_{\bf k}}\frac{\partial}{\partial a_{-\bf k}}+
\omega^{2}_{\bf k} a_{\bf k}a_{-{\bf k}}\Bigr]\Psi,
\label{SchrA1}
\end{equation}
This is the Schr\"{o}dinger equation defined in the vector space ${\bf A}_{n}$.
Re-scale the variables according to the formula 
$a_{\bf k}\rightarrow \Delta^{1/2}a_{\bf k}$ (we use the same notations for the re-scaled variables), $\Delta=const$ and consider the limiting case of 
${\bf A}_{n}$, $n\rightarrow \infty$.  
For this case using notations $\int \frac{d{\bf k}}{(2\pi)^{3}}\equiv\int \frac{d^{3}k_{i}}{(2\pi)^{3}}
=\lim_{\Delta \rightarrow 0} \Delta\sum_{k_{i}}$, $k_{i}=j\Delta $,
$j=0,\pm 1,\dots\pm n$, $n\rightarrow \infty$, we have
\begin{eqnarray}
i\dot{\Psi}
&=&
\hat{H}\Psi
\nonumber\\
&\equiv& \frac{1}{2}\int \frac{d{\bf k}}{(2\pi)^{3}}
\Bigl[
-\frac{1}{\Delta^{2}}
\frac{\partial}{\partial a_{\bf k}}
\frac{\partial}{\partial a_{-{\bf k}}}
\nonumber\\
&\;&\;\;\;\;
+
\omega^{2}_{\bf k} a_{\bf k}a_{-{\bf k}}\Bigr]\Psi,
\label{KGSch}
\end{eqnarray}
which is symbolic form of Schr\"{o}dinger equation for 3n-dimensinal, $n\rightarrow \infty$ harmonic oscillator.\footnote{Here we can formally refer to the results established in the quantum mechanics for the harmonic oscillator, however, as this is the basic case of the DI-field, we present the model in full details.}

The Hamiltonian in  
(\ref{KGSch}) can be presented as 
\begin{eqnarray}
\hat{H}
&\equiv& \frac{1}{4}\int \frac{d{\bf k}}{(2\pi)^{3}}
\Bigl[
\hat{A}_{\bf k}^{\dagger}\hat{A}_{\bf k}+
\hat{A}_{-\bf k}^{\dagger}\hat{A}_{-\bf k} 
\nonumber\\
&\;&\;\;\;\;\;\;\;
+\hat{A}_{\bf k}\hat{A}_{\bf k}^{\dagger}+
\hat{A}_{-\bf k}\hat{A}_{-\bf k}^{\dagger}
\Bigr]\nonumber\\
&\equiv& \frac{1}{2}\int \frac{d{\bf k}}{(2\pi)^{3}}
\Bigl[
\hat{A}_{\bf k}^{\dagger}\hat{A}_{\bf k}+
\hat{A}_{\bf k}\hat{A}_{\bf k}^{\dagger}
\Bigr].
\label{KGSch_op}
\end{eqnarray}
with operators $\hat{A_{\bf k}^{\dagger}}$, $\hat{A_{\bf k} }$ defined as 
\begin{eqnarray}
\hat{A_{\bf k}}^{\dagger} &=&
-\frac{1}{\Delta^{1/2}}
\frac{\partial}{\partial a_{-{\bf k}}}+
\Delta^{1/2}\omega_{\bf k}a_{\bf k}
\nonumber\\
&\equiv& -i\hat{\dot{a}}_{\bf k}+
\omega_{\bf k}\hat{a}_{\bf k},\nonumber\\
\hat{A}_{\bf k} &=& 
\frac{1}{\Delta^{1/2}}\frac{\partial}{\partial a_{\bf k}}+
\Delta^{1/2}\omega_{\bf k}a_{-{\bf k}}
\nonumber\\
&\equiv& i\hat{\dot{a}}_{-{\bf k}}+
\omega_{\bf k}\hat{a}_{-{\bf k}},
\label{op_A0}
\end{eqnarray}
$\Delta\rightarrow 0$;
with the commutator\footnote{We use
$\lim_{\Delta\rightarrow 0} \delta_{{\bf k},{\bf k}_{1}}=
(2\pi)^{3}\delta({\bf k}-{\bf k}_{1})$.}
\begin{equation}
[\hat{A}_{\bf k}, \hat{A}_{\bf k_{1}}^{\dagger}]=
2(2\pi)^{3}\delta({\bf k}-{\bf k}_{1})\omega_{\bf k},
\label{commutator_kg_0}
\end{equation}
operators $\hat{A}_{\bf k}, \hat{A}_{\bf k_{1}}^{\dagger}$ are defined in Fock space of functions $\Psi_{n,{\bf k}}$, compared with the initial Hilbert space $\Psi(a_{\bf k})$.

According to our general approach, we introduced operators 
$\hat{a}_{\bf k}=
\Delta^{1/2}a_{\bf k}$, 
$\hat{\dot{a}}_{\bf k}=i[\hat{H},\hat{a}_{\bf k}]$, so that for $\hat{\ddot{a}}_{\bf k}=i[\hat{H},\hat{\dot{a}}_{\bf k}]$ we have
\begin{equation}
\hat{\ddot{a}}_{\bf k}+\omega^{2}_{\bf k}\hat{a}_{\bf k}=0.
\label{a_oper}
\end{equation}
Now we formally define
\begin{eqnarray}
\hat{\phi}({\bf x},t) =
\int\frac{d{\bf k}}{(2\pi)^{3}}
\hat{a}_{\bf k}(t)e^{-i{\bf kx}}
\;\;\;\;\;\;
\nonumber\\ 
=
(1/2)\int\frac{d{\bf k}}{(2\pi)^{3}}
(\hat{a}_{\bf k}(t)e^{-i{\bf kx}}+
\hat{a}_{-\bf k}(t)e^{i{\bf kx}}).\nonumber\\
\;   
\label{field_oper}
\end{eqnarray}
According to (\ref{a_oper}) and (\ref{omega}), the operator $\hat{\phi}$, satisfies the equation similar to the original field equation (\ref{KG})
\begin{equation}
\Box\hat{\phi} + m^{2}\hat{\phi} = 0.
\label{KG_oper}
\end{equation}
This formally corresponds to Lagrangian presented as an operator 
\begin{equation}
\hat{\cal L}=1/2 \Bigl((\dot{\hat{\phi}})^{2}-(\nabla\hat{\phi})^{2}-
m^{2}\hat{\phi}^{2}\Bigr).
\end{equation} 
Note that if define operator of momentum 
\begin{equation}
\hat{\pi}\equiv \frac{\partial \hat{\cal L}}{\partial \dot{\hat{\phi}}}=
\dot{\hat{\phi}},
\end{equation}
we have the following commutation relations
\begin{equation}
[\hat{\phi}({\bf x}_{1}), \hat{\phi}({\bf x})]= [\hat{\pi}({\bf x}_{1}), \hat{\pi}({\bf x})]=0, 
\end{equation}
\begin{equation}
[\hat{\phi}({\bf x}_{1}), \hat{\pi}({\bf x})]=
i\delta({\bf x}-{\bf x}_{1}).
\label{commutator_phi}
\end{equation}
Consider the expression for the Hamiltonian of the operator field defined in the ordinary way, so that the expression for the operator of energy of the field will have the form  
\begin{eqnarray}
\hat{H} &=& (1/2)
\int_{V} \Bigl((\dot{\hat{\phi}})^{2}+(\nabla\hat{\phi})^{2}+
m^{2}\hat{\phi}^{2}\Bigr) d{\bf x}
\nonumber\\
&=&
\int\frac{d{\bf k}}{2(2\pi)^{3}}
\Bigl(\hat{A}^{\dagger}_{\bf k}\hat{A}_{\bf k}+
\hat{A}_{\bf k}\hat{A}^{\dagger}_{\bf k}
\Bigr),
\end{eqnarray}
which essentially has the same form as Hamiltonian in (\ref{KGSch_op}).
It may be noted that recalling (\ref{op_A0}) we can present $\hat{\phi}$ as
\begin{eqnarray}
\hat{\phi}
&=&
\frac{1}{2}
\int\frac{d{\bf k}}{(2\pi)^{3}2 \omega_{\bf k}}
\Bigl((\hat{A}_{-\bf k}+ \hat{A}_{\bf k}^{\dagger}) e^{i{\bf kx}}
\nonumber\\
&\;&\;\;\;\;
+
(\hat{A}_{\bf k}+\hat{A}^{\dagger}_{-\bf k})e^{-i{\bf kx}}\Bigr)
\nonumber\\
&=&
\int\frac{d{\bf k}}{(2\pi)^{3}2 \omega_{\bf k}}
\Bigl(\hat{A}_{\bf k}e^{-i{\bf kx}}+
\hat{A}^{\dagger}_{\bf k}e^{i{\bf kx}}\Bigr),\nonumber\\
\;
\label{def_phi_f}
\end{eqnarray}
which is the expression of operator of the field postulated within the quantum field theory. The results can be easily generalized for the case of complex Klein-Gordon field.
\\

In the similar fashion we can show that the concept of discrete interaction for the case of a field is consistent with the model of Feynman quantization.
For the real field, the equation (\ref{SchrA1}) has the same form as the equation of motion of the DI-particle, so similarly to the case of motion of a particle in space, the equation (\ref{SchrA1}) may be presented in the equivalent form of a path integral for $\alpha_{\bf k}=a_{\bf k}\Delta^{1/2}$ (here we explicitly use a small parameter $\hbar$).  
\begin{eqnarray}
\Gamma(t_{1},t_{2};{\bf k})
&\equiv& 
e^{-\frac{i}{\hbar}\int_{t_{1}}^{t_{2}}\hat{H}({\bf k})dt}\delta(t-t_{1})\delta(a-a(t_{1}))
\nonumber\\ 
&=&\int D\alpha_{\bf k} D\alpha_{-\bf k}
\exp\frac{i}{\hbar}\int_{t_{1}}^{t_{2}}
(\dot{\alpha}_{\bf k}\dot{\alpha}_{-\bf k}
\nonumber\\
&\;&\;\;\;\;\;\;\;\;\;-
(m^{2} + {\bf k}^{2}) \alpha_{\bf k}\alpha_{-\bf k})dt
\nonumber\\
&=&
\int D a_{\bf k}
D a_{-\bf k}
\exp\frac{i\Delta}{\hbar}\int_{t_{1}}^{t_{2}}
(\dot{a}_{\bf k} \dot{a}_{-\bf k} 
\nonumber\\
&\;&\;\;\;\;\;\;\;\;\;-
(m^{2} + 
{\bf k}^{2}) a_{\bf k} a_{-\bf k} )dt. 
\nonumber\\
\;
\label{KGPathInt}
\end{eqnarray}
This is the g-probability, which defines transitions between the states $a(t_{1},{\bf k})$, $a(t_{2},{\bf k})$.
In this case g-probability of transitions between the states ${\bf a}(t_{1})$, ${\bf a}(t_{2})$ is given by the formula
\begin{eqnarray}
\Gamma(t_{1},t_{2}) 
&=&
\prod_{\bf k}\Gamma(t_{1},t_{2},{\bf k})
\nonumber\\
&=&
\prod_{\bf k}\Bigl[\int D a_{\bf k}\Bigr] \exp\frac{i}{\hbar}
\int_{t_{1}}^{t_{2}}dt \int \frac{d{\bf k}}{2(2\pi)^{3}}
\nonumber\\
&\;&\;\;
(\dot{a}_{\bf k} \dot{a}_{-\bf k} 
-
(m^{2} + {\bf k}^{2})a_{\bf k}a_{-\bf k}).
\label{KGPathInt1}
\end{eqnarray}
Recalling that $\phi$ is uniquely identified by the vector $\{a_{\bf k}\}$ in  {\bf A}, we use
\[
\prod_{\bf k}\Bigl[\int D a_{\bf k}\Bigr]= \int D\phi.
\]
after integration in (\ref{KGPathInt1}) by ${\bf k}$ we obtain 
\begin{equation}
\Gamma(t_{1},t_{2}) = \int D\phi\exp\frac{i}{\hbar}\int_{t_{1}}^{t_{2}}
{\cal L}(\phi,t)d{\bf x}dt.
\label{Prop_KG1}
\end{equation} 
here 
\begin{equation}
{\cal L} = 1/2\Bigl[(\dot{\phi})^{2} - 
(\nabla\phi)^{2} + m^{2}\phi^{2}\Bigr],  
\end{equation}
$\Gamma$ defines the g-probability of transition between the states
$\phi(t_{1})$, $\phi(t_{2})$.
That is the path integral defined in the space of functions 
$\phi$ which represent distribution of the field in $E^{3}$. 

It may be noted that similarly to the case of the particle, in limit $\hbar\rightarrow 0$, the expression (\ref{Prop_KG1}) leads to the principle of the least action and subsequently to the Klein-Gordon equation for the classical field $\phi$. This confirms the correspondence between the equations for the classical and DI-fields. 

We considered the model of the free DI-field, as related to the linear classical equation, based on the definition of the elementary states of the field as amplitudes of harmonics.
Recalling the condition of orthogonality of the harmonics, the behavior of the system was presented as motion in $E^{n}$, $n\rightarrow \infty$.
Strictly speaking we also should verify that the functions which represent g-probabilities of the different sates are orthogonal,
\begin{equation}
\sum_{n=0}^{\infty}\Psi_{E_{n}}(a_{i})\Psi_{E_{n}}(a_{j})=\delta_{i,j};
\label{orthogonality}
\end{equation}  
the expression is true in the case of oscillator, where $E_{n}\equiv E_{n}(m)$ is the energy of an oscillator, defined for the particular value of the parameter $m$. 
Recalling that the classical field can be obtained as a limiting case of the DI-field, we conclude that the DI-field, defined as above, proves to exist for the classical field with any value of the parameter $m$.
\\
\\
{\em Maxwell field.}
\\

Consider $A_{\mu}=\{A_{0},{\bf A}\}$, so that 
\begin{eqnarray}
\Box A_{\mu}=0,\nonumber\\
\nabla {\bf A}=0,\;\; A_{0}=0.
\label{maxwell}
\end{eqnarray}
For
\begin{equation}
A_{i}(x,t)=
\int \frac{d{\bf k}}{(2\pi)^{3}}a_{i;\bf k}(t) e^{-i{\bf kx}},
\end{equation}
$i=1,2,3$, the system (\ref{maxwell}) is presented as 
\begin{eqnarray}
\ddot{a}_{i,{\bf k}}=k_{0}^{2}a_{i,{\bf k}},\;\;\;
k_{i}a_{i,{\bf k}}=0,
\end{eqnarray}
for each value of the parameter ${\bf k}$ the state of the system is described by two independent variables. 
For the given ${\bf k}$ these values may be defined as
\begin{eqnarray}
b_{1}=({\bf a}\cdotp{\bf e}_{1}),\;\;\;
b_{2}=({\bf a}\cdotp{\bf e}_{2}),
\end{eqnarray}
where $({\bf e}_{1}\cdotp{\bf k})=({\bf e}_{2}\cdotp{\bf k})=
({\bf e}_{1}\cdotp{\bf e}_{2})=0$,
$\|{\bf e}_{1}\|=\|{\bf e}_{2}\|=1$, so that 
${\bf a}=b_{1}{\bf e}_{1}+b_{2}{\bf e}_{2}$.
Consequently we have
\begin{eqnarray}
\ddot{b}_{i,{\bf k}}=k_{0}^{2}b_{i,{\bf k}},\;\;i=1,2.
\end{eqnarray}
According to our general approach, assume that the set of values of the components of the vector ${\bf b}_{1,2}$ represents the particular elementary state. For the DI-system defined in the vector space 
${\bf B}\equiv \{{\bf B}_{1}\oplus\bf{B}_{2}\}$ we can also define g-probability
$\Psi({\bf b}_{1}, {\bf b}_{2})$ of entering the state specified by the set of parameters ${\bf b}_{1}, {\bf b}_{2}$. Recalling that ${\bf b}_{1}$, ${\bf b}_{2}$ belong to orthogonal subspaces in $\{{\bf B}_{1}\oplus\bf{B}_{2}\}$, using the same assumptions about topology of ${\bf B}$ and particle-like behavior of the system, we  establish 
\begin{eqnarray}
i\dot{\Psi}&=&\hat{H}\Psi
\nonumber\\
&\equiv&
\frac{1}{2}\int \frac{d{\bf k}}{(2\pi)^{3}}
\Bigl[-\frac{1}{\Delta^{2}}\Bigl(\frac{\partial}{\partial b_{1,{\bf k}}}
\frac{\partial}{\partial b_{1,-{\bf k}}}
+
\nonumber\\
&\;&
\frac{\partial}{\partial b_{2,{\bf k}}}
\frac{\partial}{\partial b_{2,-{\bf k}}}
\Bigr)
\nonumber\\
&\;&+
k_{0} (b_{1,{\bf k}}b_{1,-{\bf k}}+b_{2,{\bf k}}b_{2,-{\bf k}})\Bigr]\Psi,
\nonumber\\
\;
\label{Schr_mak}
\end{eqnarray}
with the meaning $\int \frac{d{\bf k}}{(2\pi)^{3}}=\lim_{\Delta\rightarrow 0}\Delta\sum_{k_{i}}$, $\Delta=const$.
According to our general approach, we define operators 
$\hat{b}_{i,{\bf k}}=\Delta^{1/2}b_{i,{\bf k}}$, 
\\
so for $\hat{\ddot{b}}_{i,{\bf k}}=-[\hat{H},[\hat{H},\hat{b}_{i,{\bf k}}]]$ we have  
\begin{equation}
\hat{\ddot{b}}_{i,{\bf k}} + k_{0}\hat{b}_{i,{\bf k}} = 0.
\label{A_oper_mak}
\end{equation}
Consider 
$\Psi=\prod_{{\bf k},i=1,2}\psi_{i; {\bf k}}(b_{i,{\bf k}}, b_{i,{\bf k}})$. 
For the stationary states $\psi_{i;{\bf k},n}$, with the energy  $\epsilon_{n}=nk_{0}$, substitution into (\ref{Schr_mak}) gives
\begin{equation}
nk_{0}\psi_{i;{\bf k},n}=
1/4\Bigl(\hat{B}_{i;{\bf k}}^{\dagger}\hat{B}_{i;{\bf k}}+
\hat{B}_{i;{-\bf k}}^{\dagger}\hat{B}_{i;{-\bf k}}
+ 2k_{0}\Bigr)\psi_{i;{\bf k},n}.
\end{equation}
where operators $\hat{B}_{i;k}^{\dagger}$, $\hat{B}_{i;k}$ are defined as 
\begin{eqnarray}
\hat{B}_{i;{\bf k}}^{\dagger} &=& 
-\frac{1}{\Delta^{1/2}}\frac{\partial}{\partial b_{i;{-\bf k}}}+
\Delta^{1/2}k_{0}b_{i;\bf k}
\nonumber\\
&\;&\;\;
\equiv 
-i\hat{\dot{b}}_{i;\bf k}+
k_{0}\hat{b}_{i;\bf k},\nonumber\\
\hat{B}_{i;\bf k} &=& 
\frac{1}{\Delta^{1/2}}\frac{\partial}{\partial b_{i;\bf k}}+
\Delta^{1/2}k_{0}b_{i;{-\bf k}}
\nonumber\\
&\;&\;\;
\equiv 
i\hat{\dot{b}}_{i;-\bf k}+
k_{0}\hat{b}_{i;-\bf k};
\label{op_mak}
\end{eqnarray}
with the commutator ($\Delta\rightarrow 0$). 
\begin{equation}
[\hat{B}_{i;\bf k}^{\dagger}\hat{B}_{j;{\bf k}_{1}}]=
2(2\pi)^{3}\delta({\bf k}-{\bf k}_{1})\delta_{i;j}k_{0}.
\label{commutator_mak}
\end{equation}
Now we can formally introduce an operator\footnote{Similarly to $\hat{\bf r}$ in the case of a particle, the vector 
$\hat{\bf A}\equiv\{\hat{A}_{1}, \hat{A}_{2}, \hat{A}_{3}\}$ means a set of operators $\hat{A}_{i}$, so that each acts on the particular term of the function $\Psi({\bf A})\equiv\Psi(A_{1},A_{2},A_{3})$.}
\begin{equation}
\hat{\bf A}({\bf x},t)=\int \frac{d{\bf k}}{(2\pi)^{3}}
\Bigl({\bf e}_{1}\hat{b}_{1;{\bf k}}(t)e^{i{\bf kx}}
+{\bf e}_{2}\hat{b}_{2;{\bf k}}(t)e^{i{\bf kx}}
\Bigr),   
\label{field_oper_mak}
\end{equation}
According to (\ref{A_oper_mak}), operator $\hat{A}({\bf x},t)$ satisfies (\ref{maxwell})
\begin{eqnarray}
\partial_{t}^{2}\hat{A}_{i} - \Delta\hat{A}_{i} = 0,\nonumber\\
\nabla \hat{A}_{i}=0,\;\; \hat{A}_{0}=0.
\label{mak_oper}
\end{eqnarray}
This corresponds to classical Maxwell 
\\
Lagrangian presented as an operator 
\begin{equation}
\hat{\cal L}=(1/2) (\partial_{\nu}{\hat{A}}_{\mu}-
\partial_{\mu}{\hat{A}}_{\nu})
(\partial^{\nu}{\hat{A}}^{\mu}-
\partial^{\mu}{\hat{A}}^{\nu});
\end{equation} 
consequently for the operator of momentum we have
\begin{equation}
\hat{\pi}^{i}\equiv \frac{\partial \hat{L}}{\partial \dot{\hat{A}}}_{i}=
-\dot{\hat{A^{i}}}.
\end{equation}
The definition of the operators implies the following commutation relations
\begin{equation}
[\hat{A}_{i}({\bf x}_{1}), \hat{A}_{j}({\bf x})]= 
[\hat{\pi}_{i}({\bf x}_{1}), \hat{\pi}_{j}({\bf x})]=0,
\end{equation}
\begin{equation}
[\hat{A}_{i}({\bf x}_{1}), \hat{\pi}_{j}({\bf x})]=
i\Bigl(\delta_{i,j}+\frac{\partial_{i}\partial_{j}}
{\nabla^{2}}\Bigr)\delta({\bf x}-{\bf x}_{1}),
\label{commutator_phi_mak}
\end{equation}
Note that, using the definition (\ref{op_mak}) the expression (\ref{field_oper_mak}) may be presented in the form
\begin{equation}
\hat{\bf A}=\int\frac{d{\bf k}}{(2\pi)^{3}2 k_{0}}\sum_{i=1}^{2}{\bf e}_{i}\Bigl(\hat{B}_{i}e^{-i{\bf kx}}+\hat{B_{i}^{\dagger}}e^{i{\bf kx}}\Bigr),
\label{def_phi_mak}
\end{equation}
which is the conventional form of the operator of electro-magnetic field postulated in the quantum field theory. 
\\
\\
{\em Massive vector field.}
\\

Consider massive real vector field $A_{i}$, which is described by Proka equation.  
\begin{eqnarray}
\Box A_{i} +m^{2}A_{i}=0,\nonumber\\
\partial_{i} A^{i}=0, 
\label{proka}
\end{eqnarray}
Similarly to the previous case consider
\begin{equation}
a_{i;{\bf k}}=\int_{V}A_{i}(x,t)e^{i{\bf kx}}d{\bf x},
\end{equation}
$i=1,2,3,4$, so that the system (\ref{proka}) is presented as 
\begin{eqnarray}
\ddot{a}_{i,{\bf k}}=\omega_{\bf k}^{2}a_{i,{\bf k}}\nonumber\\
k_{i}a_{i,{\bf k}}=0,
\end{eqnarray}
here $\omega_{\bf k}^{2}={\bf k}^{2}+m^{2}$.
We use $a_{0;{\bf k}}=0$, so that for each value of the parameter ${\bf k}$ the state of the system is described by the components of the vector 
${\bf a}_{\bf k}=a_{i;{\bf k}}$, $i=1,2,3$. 

According to our general approach we assume that the set of values of the components of the vector 
${\bf a}_{\bf k}=a_{1,{\bf k}}{\bf e}_{1}+a_{2,{\bf k}}{\bf e}_{2}+
a_{3,{\bf k}}{\bf e}_{3}$ 
represents the particular elementary state. For the DI-field defined in the vector space 
${\bf A}\equiv \{{\bf A}_{1}\oplus{\bf A}_{2}\oplus{\bf A}_{3}\}$
\\
 we define g-probability
$\Psi(a_{1,{\bf k}}, a_{2,{\bf k}}, a_{3,{\bf k}})$. 
Repeating all the previous arguments, we establish 
\begin{eqnarray}
i\dot{\Psi} &=& \hat{H}\Psi
\nonumber\\
&\equiv&
\frac{1}{2}\int \frac{d{\bf k}}{(2\pi)^{3}}
\sum_{i=1}^{3}\Bigl[-\frac{1}{\Delta^{2}}
\frac{\partial}{\partial a_{i,{\bf k}}}
\frac{\partial}{\partial a_{i,{-\bf k}}} 
\nonumber\\
&\;&\;\;\;\;\;\;+ 
\omega_{\bf k}^{2} a_{i,{\bf k}} a_{i,{-\bf k}} 
\Bigr]\Psi,
\label{Schr_proka}
\end{eqnarray}
Similarly to the previous case, we introduce operators 
$\hat{a}_{i,{\bf k}}=\Delta^{1/2}a_{i,{\bf k}}$, so for 
\\
$\hat{\ddot{a}}_{i,{\bf k}}=-[\hat{H},[\hat{H},\hat{a}_{i,{\bf k}}]]$   
we have
\begin{equation}
\hat{\ddot{a}}_{i,{\bf k}} + \omega_{\bf k}^{2}\hat{a}_{i,{\bf k}} = 0.
\label{A_oper_proka}
\end{equation}
Consider 
$\Psi=\prod_{{\bf k},i=1,2,3}\psi_{i; {\bf k}}(a_{i,{\bf k}}, a_{i,{-\bf k}})$. 
For the stationary states $\psi_{i;{\bf k},n}$, with the energy  $\epsilon_{n}=n\omega_{\bf k}$, substitution into (\ref{Schr_proka}) gives
\begin{equation}
n\omega_{\bf k}\psi_{i;{\bf k},n}=
\frac{1}{2}\Bigl(\hat{P}_{i;{\bf k}}^{\dagger}\hat{P}_{i;{\bf k}}
+\hat{P}_{i;{-\bf k}}^{\dagger}\hat{P}_{i;{-\bf k}}
+2\omega_{\bf k}\Bigr)\psi_{i;{\bf k},n};
\end{equation}
operators $\hat{P}_{i;k}^{\dagger}$, $\hat{P}_{i;k}$ are defined as 
\begin{eqnarray}
\hat{P}_{i;{\bf k}} &=& 
-\frac{1}{\Delta^{1/2}}\frac{\partial}{\partial a_{i;{-\bf k}}}+
\Delta^{1/2}\omega_{\bf k}a_{i;\bf k}
\nonumber\\
&\equiv& 
-i\hat{\dot{a}}_{i;\bf k}+
\omega_{\bf k}\hat{a}_{i;\bf k},\nonumber\\
\hat{P}_{i;\bf k}^{\dagger} &=& 
\frac{1}{\Delta^{1/2}}\frac{\partial}
{\partial a_{i;\bf k}}+
\Delta^{1/2}\omega_{\bf k}a_{i;{-\bf k}}
\nonumber\\
&\equiv& 
i\hat{\dot{a}}_{i;-\bf k}+
\omega_{\bf k}\hat{a}_{i;-\bf k};
\label{op_proka}
\end{eqnarray}
with the commutator ($\Delta\rightarrow 0$) 
\begin{equation}
[\hat{P}_{i;\bf k}^{\dagger}\hat{P}_{j;\bf k_{1}}]=
2(2\pi)^{3}\delta({\bf k}-{\bf k_{1}})\delta_{i;j}\omega_{\bf k}.
\label{commutator_proka}
\end{equation}
Now define
\begin{equation}
\hat{A}({\bf x},t)=\int \frac{d{\bf k}}{(2\pi)^{3}} 
\sum_{j=1}^{3}{\bf e}_{j}\hat{a}_{j,{\bf k}}(t)e^{i{\bf kx}}, 
\label{field_oper_proka}
\end{equation}
which, according to (\ref{A_oper_proka}), satisfies (\ref{proka})
\begin{eqnarray}
\partial_{t}^{2}\hat{A}_{j} - \Delta\hat{A}_{j} + 
m^{2}\hat{A}_{j} = 0,\nonumber\\
\partial_{j} A^{j}=0. 
\label{KG_oper_proka}
\end{eqnarray}
This corresponds to the operator of Lagrangian 
\begin{eqnarray}
\hat{\cal L}=\Bigl(-(1/4)(\partial_{\nu}{\hat{A}}_{\mu}-
\partial_{\mu}{\hat{A}}_{\nu})
(\partial^{\nu}{\hat{A}}^{\mu}-
\partial^{\mu}{\hat{A}}^{\nu})
\nonumber\\
+
(1/2)m^{2}{\hat{A}}_{\mu}{\hat{A}}^{\mu} 
\Bigr).
\nonumber\\
\;
\end{eqnarray} 
For operator of momentum 
\begin{equation}
\hat{\pi}^{i}\equiv \frac{\partial \hat{L}}{\partial \dot{\hat{A}}}_{i}=
-\dot{\hat{A^{i}}},
\end{equation}
we have the following commutation relations
\begin{equation}
[\hat{A}_{i}({\bf x}_{1}), \hat{A}_{j}({\bf x})]= 
[\hat{\pi}_{i}({\bf x}_{1}), \hat{\pi}_{j}({\bf x})]=0,
\end{equation}
\begin{equation}
[\hat{A}_{i}({\bf x}_{1}), \hat{\pi}_{j}({\bf x})]=
\delta_{i,j}\delta({\bf x}-{\bf x}_{1}).
\label{commutator_phi_proka}
\end{equation}
Using the formulae (\ref{op_proka}) the expression (\ref{field_oper_proka}) may be presented in as
\begin{equation}
\hat{\bf A}=\int\frac{d{\bf k}}{(2\pi)^{3}2 \omega_{\bf k}}\sum_{i=1}^{3}{\bf e}_{i}\Bigl(\hat{A}_{i}e^{-i{\bf kx}}+\hat{A_{i}^{\dagger}}e^{i{\bf kx}}\Bigr),
\label{def_phi_proka}
\end{equation}
which is the conventional form of the operator defined in the quantum field theory.
\\

The operator formalism established for the DI-fields with an integer spin is similar to the formalism established within the quantum field theory. 
Now consider the model of bi-spinor DI-field. Here we use the property that for a pair of bi-spinors the inner product can be defined as 
\begin{equation}
\eta_{a}\cdotp\eta_{b}=\eta_{a}^{1}\eta_{b}^{2}-\eta_{a}^{2}\eta_{b}^{1}+
\eta_{a}^{3}\eta_{b}^{4}-\eta_{a}^{4}\eta_{b}^{3};
\label{gras_dot_prod}
\end{equation}
this implies anti-commutation relation
\begin{equation}
\eta_{a}\cdotp\eta_{b}=-\eta_{b}\cdotp\eta_{a}.
\label{eta_anticommuattion}
\end{equation}

\subsection{\rm Dirak field}
The bi-spinor field $\eta=(\eta_{1}, \eta_{2}, \eta_{3}, \eta_{4})$
is defined by the set of equations
\begin{equation}
i\gamma_{0}\partial_{t}\eta = -i\hat{\partial_{n}}\eta + m\eta,
\label{dirak}
\end{equation}
$m$ is a nonnegative constant, 
$\hat{\partial_{n}}=\gamma_{n}\partial_{x_{n}}$;
$\gamma_{0}, \gamma_{n}$, $n=1,2,3$ are Dirak matrixes.

The plane wave type solutions of (\ref{dirak}) may be presented in the form 
\begin{eqnarray}
\eta &=& \sum_{i=1,2}\int \frac{d{\bf k}}{(2\pi)^{3}}
\Bigl( 
h_{i,{\bf k}}(t)u_{i}(k)e^{-i{\bf kx}}
\nonumber\\
&\;&\;\;\;\;
+
g_{i,{\bf k}}(t)v_{i}(k)e^{i{\bf kx}}
\Bigl),
\label{eta1}
\end{eqnarray} 
here $h_{i,{\bf k}}(t)=h(i,{\bf k})e^{k_{0}t}$, 
$g(i,{\bf k})e^{-k_{0}t}$ are the `amplitueds' of the harmonics; $u_{i}(k_{0},{\bf k})$, $v_{i}(k_{0},{\bf k})$ basis bi-spinors. 
The spinors $u_{i}(k)$, $v_{i}(k)$ constitute a basis of the vector space, with an arbitrary vector presented in the form (\ref{eta1}). This vector space allows inner product, which according to (\ref{gras_dot_prod}) is defined as
\begin{equation}
\eta_{a}\cdotp\eta_{b}\equiv\sum_{i=1,2;{\bf k}}\eta_{a;i,{\bf k}}
\eta_{b; i,{\bf k}}= h_{a;i,{\bf k}} h_{b;i,{\bf k}} +
g_{a;i,{\bf k}} g_{b;i,{\bf k}};
\end{equation}
note that anti-commutation relation (\ref{eta_anticommuattion}) will require anti-commutation relation defined for  the variables 
$h_{a;i,{\bf k}}, g_{a;i,{\bf k}}$.
In this case an algebraic function defined on the set of spinors may be identically defined on the set of amplitudes
$\Phi(\eta)\equiv\Phi(h_{i,{\bf k}}, g_{i,{\bf k}})$, where
$h_{i,{\bf k}}, g_{i,{\bf k}}$ are Grassmann variables;
we also require $h_{i,{\bf k}}g_{i,{\bf k}}=0$.

We define a set of Grassmann variables $d_{i, {\bf k}}\equiv d(i, {\bf k};t)$, according to the formulae
\begin{eqnarray}
d_{i, {\bf k}}&=&(h_{i, {\bf k}}-g_{i, {\bf k}})/(2\omega_{\bf k}), 
\nonumber\\
\dot{d}_{i, {\bf k}}&=&(h_{i, {\bf k}}+g_{i, {\bf k}})/(2i),
\end{eqnarray}
so that 
\begin{eqnarray}
h_{i, {\bf k}}&=&i\dot{d}_{i, {\bf k}}+ \omega_{\bf k} d_{i, {\bf k}}, 
\nonumber\\
g_{i, {\bf k}}&=&i\dot{d}_{i, {\bf k}}- \omega_{\bf k} d_{i, {\bf k}},
\label{dirak_d}
\end{eqnarray}
This implies
\begin{equation}
\ddot{d}_{i,{\bf k}}=-\omega^{2}_{\bf k} d_{i;{\bf k}}.
\label{b_motion}
\end{equation}
For each $d_{i;{\bf k}}$ we define a conjugate 
$d_{i;{\bf k}}^{*} = d_{i;{-\bf k}}$, so that we also have
\begin{equation}
\ddot{d}_{i,{\bf k}}^{*}=-\omega^{2}_{\bf k} d_{i;{\bf k}}^{*}.
\label{b_motion_1}
\end{equation}
The set of 4n-dimensional Grassmann variables 
${\bf d}=\{d_{i,{\bf k}},d_{i,{\bf k}}^{*}\}$ 
constitute a vector space ${\bf D}$
with the inner product 
\[
({\bf d}_{a}\cdotp{\bf d}_{b})=
\sum_{i=1,2,{\bf k}}( d_{i_{a};{\bf k}_{a}}\cdotp d_{i_{b};{\bf k}_{b}} +
d_{i_{a};{\bf k}_{a}}^{*}\cdotp d_{i_{b};{\bf k}_{b}}^{*} ).
\]

We suppose that the elementary state of the DI-field, which corresponds to the classical Dirak field, is an $n$-dimensional, $n\rightarrow \infty$ Grassmann vector ${\bf d}$ defined as a set of amplitudes $d_{i;{\bf k}}$, so that behavior of the system may be presented as motion in the Grassmann vector space ${\bf D}=\{{\bf D}\oplus{\bf D}^{*}\}$. For each particular moment in time we define g-probability $\Psi({\bf d})$ of entering or transition between elementary states defined in the vector space  ${\bf D}$, so that the behavior of g-probability $\Psi$ in time which corresponds to (\ref{b_motion}), (\ref{b_motion_1}) will be defined by equation of the type (\ref{gras_Schr}). We suppose that in our case the function $\theta$ in (\ref{gras_Schr}) has the form  
\begin{eqnarray}
\theta({\bf d},{\bf d}_{1}) &=&
\sum_{i=1,2}\int d{\bf k}\Bigl[1+
\omega_{\bf k}^{2}d_{i,\bf k}^{*}d_{i,\bf k}
d_{1;i,{\bf k}}^{*}d_{1;i,{\bf k}}
\Bigr]
\nonumber\\
&\;&\;\;\times 
\prod_{{\bf p}\neq {\bf k}}
(d_{i,{\bf p}}^{*}+d_{1;i,{\bf p}}^{*})
(d_{i,{\bf p}}+d_{1;i,{\bf p}})
\nonumber\\
\;
\end{eqnarray}
so that the equation of motion of the system, after appropriate re-scaling of variables, is defined as
\begin{eqnarray}
\dot\Psi &=& \hat{H}\Psi
\nonumber\\
&\equiv &
\frac{1}{2}\int d{\bf k}\Bigl[
\frac{1}{\Delta^{2}}\Bigl(
\frac{\partial}{\partial d_{1,{\bf k}}}
\frac{\partial}{\partial d_{1,{\bf k}}^{*}}+
\frac{\partial}{\partial d_{2,{\bf k}}}
\frac{\partial}{\partial d_{2,{\bf k}}^{*}}\Bigr)
\nonumber\\
&\;&\;\;\;+ 
\omega_{\bf k}^{2}(d_{1,{\bf k}}^{*} d_{1,{\bf k}}+
d_{2,{\bf k}}^{*} d_{2,{\bf k}})
\Bigr]\Psi.
\label{dirak_schr_1}
\end{eqnarray}
Consider $\Psi=\prod_{i=1,2;\bf k}\psi_{i,\bf k}(d_{i,\bf k})$.
For the stationary states $\psi_{i,{\bf k},n,m}(a_{\bf k})$ with the values of energy $\epsilon_{{\bf k};n,m}=(n+m)\omega_{\bf k}$ the formula (\ref{dirak_schr_1}) can be presented as
\begin{equation}
(n+m)\omega_{\bf k}\psi_{i,{\bf k},n,m}=
(\hat{F}_{i,{\bf k}}^{\dagger}\hat{F}_{i,{\bf k}}-
\hat{E}_{i,{\bf k}}\hat{E}^{\dagger}_{i,{\bf k}})
\psi_{i,{\bf k},n,m},
\label{dirak_schr}
\end{equation}
$\psi_{i,{\bf k},n,m}$ is defined so that the operators 
$\hat{E}_{i,{\bf k}}, \hat{E}^{\dagger}_{i,{\bf k}}$ 
are `changing' the value of $n$, and
$\hat{F}_{i,{\bf k}}, \hat{F}^{\dagger}_{i,{\bf k}}$ the value of $m$;
operators $\hat{F}_{i,{\bf k}}$, $\hat{F}_{i,{\bf k}}^{\dagger}$, 
$\hat{E}_{i,{\bf k}}$, $\hat{E}_{i,{\bf k}}^{\dagger}$
are defined as 
\begin{eqnarray}
\hat{F}_{i,{\bf k}} &=& 
\frac{1}{\Delta^{1/2}}\frac{\partial}{\partial d_{i,{\bf k}}^{*}}+
\Delta^{1/2}\omega_{\bf k}d_{i,{\bf k}}
\nonumber\\
&\;&\;\;\;
\equiv
i\hat{\dot d}_{i,{\bf k}}+\omega_{\bf k}\hat{d}_{i,{\bf k}},\nonumber\\
\hat{F}_{i,{\bf k}}^{\dagger} &=& 
\frac{1}{\Delta^{1/2}}
\frac{\partial}{\partial d_{i,{\bf k}}}+
\Delta^{1/2}\omega_{\bf k}d_{i,{\bf k}^{*}}
\nonumber\\
&\;&\;\;\;
\equiv
-i\hat{\dot d}_{i,{\bf k}}^{*}+\omega_{\bf k}\hat{d}_{i,{\bf k}}^{*},
\label{oper_F}
\end{eqnarray}
and (operators $\hat{E}_{i,{\bf k}}$, 
$\hat{E}_{i,{\bf k}}^{\dagger}$ are {\em defined} as conjugate)
\begin{eqnarray}
\hat{E}_{i,{\bf k}}^{\dagger} &=& 
\Bigl(\frac{1}{\Delta^{1/2}}\frac{\partial}{\partial d_{i,{\bf k}}^{*}}-
\Delta^{1/2}\omega_{\bf k}d_{i,{\bf k}}\Bigr)
\nonumber\\
&\equiv &
i\hat{\dot d}_{i,{\bf k}}-\omega_{k}\hat{d}_{i,{\bf k}},\nonumber\\
\hat{E}_{i,{\bf k}} &=& 
-\Bigl(\frac{1}{\Delta^{1/2}}\frac{\partial}{\partial d_{i,{\bf k}}}-
\Delta^{1/2}\omega_{\bf k}d_{i,{\bf k}}^{*}\Bigr)
\nonumber\\
&\equiv &
i\hat{\dot d}_{i,{\bf k}}^{*}+\omega_{\bf k}\hat{d}_{i,{\bf k}}^{*},
\label{oper_E}
\end{eqnarray}
with anti-commutation relations ($\Delta\rightarrow 0$)
\begin{eqnarray}
\{\hat{F}_{i,{\bf k}}^{\dagger}\hat{F}_{j,{\bf k}_{1}}\} &=&
\{\hat{E}_{i,{\bf k}}^{\dagger}\hat{E}_{j,{\bf k}_{1}}\}
\nonumber\\
&=&
2(2\pi)^{3}\delta({\bf k}-{\bf k}_{1})\delta_{i,j}\omega_{\bf k},
\nonumber\\
\;
\label{anticommutator}
\end{eqnarray}
and
\begin{eqnarray}
\{\hat{F}_{i,{\bf k}}^{\dagger}\hat{F}_{j,{\bf k}_{1}}^{\dagger}\}=
\{\hat{F}_{i,{\bf k}}\hat{F}_{j,{\bf k}_{1}}\}=0,
\nonumber\\
\{\hat{E}_{i,{\bf k}}^{\dagger}\hat{E}_{j,{\bf k}_{1}}^{\dagger}\}=
\{\hat{E}_{i,{\bf k}}\hat{E}_{j,{\bf k}_{1}}\}=0.
\label{anticommutator_1}
\end{eqnarray}
We defined 
$\hat{d}_{i,{\bf k}}=\Delta^{1/2}d_{i,{\bf k}}$,
$\hat{d^{*}}_{i,{\bf k}}=\Delta^{1/2}d^{*}_{i,{\bf k}}$ so for 
$\hat{\ddot{d}}_{i,{\bf k}}=-[\hat{H},[\hat{H},\hat{d}_{i,{\bf k}}]]$
and
\\
$\hat{\ddot{d^{*}}}_{i,{\bf k}}=-[\hat{H},[\hat{H},\hat{d^{*}}_{i,{\bf k}}]]$
we have
\begin{eqnarray}
\hat{\ddot{d}}_{i,{\bf k}} + \omega_{\bf k}^{2}\hat{d}_{i,{\bf k}} = 0, \;\;\;
\hat{\ddot{d^{*}}}_{i,{\bf k}} + \omega_{\bf k}^{2}\hat{d^{*}}_{i,{\bf k}} = 0.
\label{A_oper_dirak}
\end{eqnarray}
The similar equations may be specified for 
\[
\hat{h}_{i,{\bf k}}=i\hat{\dot{d}}_{i,{\bf k}}+ \omega\hat{d}_{i,{\bf k}}\equiv
\hat{F}_{i,{\bf k}}
\]
and
\[
\hat{g}_{i,{\bf k}}=i\hat{\dot{d}}_{i,{\bf k}}- \omega\hat{d}_{i,{\bf k}}\equiv
\hat{E}_{i, {\bf k} }^{\dagger}.
\]
According to (\ref{oper_F}), (\ref{oper_E}) we have
\begin{eqnarray}
\dot{\hat{F}}_{i,{\bf k}} =-i\omega_{\bf k}\hat{F}_{i,{\bf k}}, \;\;
\dot{\hat{F}}^{\dagger}_{i,{\bf k}} 
=i\omega_{\bf k}\hat{F}^{\dagger}_{i,{\bf k}},\nonumber\\
\dot{\hat{E}}_{i,{\bf k}} 
=-i\omega_{\bf k}\hat{E}_{i,{\bf k}}, \;\;
\dot{\hat{E}}^{\dagger}_{i,{\bf k}} 
=i\omega_{\bf k}\hat{E}^{\dagger}_{i,{\bf k}}.
\label{oper_time}
\end{eqnarray}

Now, similarly to the case of a vector field we introduce bi-spinor operators
$\hat{\eta}=(\hat{\eta}_{1},\hat{\eta}_{2})$
\begin{eqnarray}
\hat{\eta}_{i}({\bf x},t)&=&
\int \frac{d{\bf k}}{(2\pi)^{3}} 
\Bigl( 
u_{i}\hat{h}_{i,{\bf k}} 
e^{-i{\bf kx}}+
v_{i}\hat{g}_{i,{\bf k}} 
e^{i{\bf kx}}
\Bigr)
\nonumber\\
&\equiv&
\int \frac{d{\bf k}}{(2\pi)^{3}} 
\Bigl( 
u_{i}\hat{F}_{i,{\bf k}} 
e^{-i{\bf kx}}+
v_{i}\hat{E}_{i,{\bf k}}^{\dagger}  
e^{i{\bf kx}}
\Bigr),
\nonumber\\
\;
\label{field_oper_dirak}
\end{eqnarray}
which, according to (\ref{eta1}), (\ref{oper_time}) satisfy the equation similar to the original field equation (\ref{dirak})
\begin{eqnarray}
i\gamma_{0}\partial_{t}\hat{\eta} = -i\hat{\partial_{n}}\hat{\eta} + m\hat{\eta}.
\label{dirak_oper}
\end{eqnarray}
We also define the conjugate operator according to the formula
\begin{eqnarray}
\hat{\eta}_{i}^{\dagger} ({\bf x},t)=
\int \frac{d{\bf k}}{(2\pi)^{3}} 
\Bigl( 
\bar{u}_{i}\hat{F}_{i,{\bf k}}^{\dagger} 
e^{i{\bf kx}}+
\bar{v}_{i}\hat{E}_{i,{\bf k}} 
e^{-i{\bf kx}}
\Bigr).
\nonumber\\
\;
\label{field_oper_dirac_c}
\end{eqnarray}
The equation (\ref{dirak_oper}) corresponds to operator of Lagrangian 
\begin{equation}
\hat{\cal L}=\frac{i}{2}
\Bigl(
\dot{\bar{\eta}}\gamma^{\mu}(\partial_{\mu}\hat{\eta})-
(\partial_{\mu}\bar{\eta})
\gamma^{\mu}\hat{\eta}-
2m\bar{\eta}\hat{\eta}
\Bigr)
,
\end{equation} 
$\bar{\eta}=\hat{\eta}^{\dagger}\gamma^{0}$,
operators in the brackets modify the functions on their left.
If define operator of momentum
\begin{equation}
\hat{\pi}\equiv \frac{\partial \hat{L}}{\partial \dot{\hat{\eta}}}=
i\hat{\eta^{\dagger}},
\end{equation}
we can specify
\begin{eqnarray}
\{\hat{\eta}({\bf x}_{1}), \hat{\eta}({\bf x})\}=
\{\hat{\pi}({\bf x}_{1}), \hat{\pi}({\bf x})\}=0,\nonumber\\
\{\hat{\eta}({\bf x}_{1}), \hat{\pi}({\bf x})\}=
i\delta({\bf x}-{\bf x}_{1}).
\label{commutator_eta}
\end{eqnarray}
The expression for the Hamiltonian defined in the ordinary way, implies that \begin{eqnarray}
\hat{H} &=& 
i\int_{V} (\hat{\eta}^{\dagger}\dot{\hat{\eta}}-
\dot{\hat{\eta}}^{\dagger}\hat{\eta}) d{\bf x} 
\nonumber\\
&\equiv &
\sum_{i=1,2} \int \frac{d{\bf k}}{(2\pi)^{3}}
(\hat{F}_{i,{\bf k}}^{\dagger}\hat{F}_{i,{\bf k}}-
\hat{E}_{i,{\bf k}}\hat{E}^{\dagger}_{i,{\bf k}}).
\nonumber\\
\;
\end{eqnarray}
The results are similar to the mathematical formalism of secondary quantization established for Dirak field. 
\\

It may be noted that the model of the behavior of the DI-fields is similar to the model of the quantum fields established in the conventional quantum field theory. This may be attributed to the fact that the equations of the field are closely related to models of oscillators of the particular types. Within certain, more intuitive approaches of the conventional quantum field theory, at least for the scalar case, the field is postulated to represent a set of oscillators. 
The present approach demonstrates that as long as the discrete character of interaction is postulated, the Schr\"{o}dinger equation for the amplitude of the field, which is established provided the condition of the type (\ref{4}), appears to be the equation for an oscillator, with all the particular results to follow. 
\\
\\ 

\section{\rm Comparison with the conventional formulation of principles of quantum mechanics}
The conventional formulation of quantum mechanics is based on of mathematical apparatus, which describes the behavior of observables. 
The quantitative description of behavior of the quantum system is based on the following formalism (in brief) (see for example A. Bohm \cite{abohm}, or Landau and Lifshitz \cite{LL}).
 
1. A physical observable, is represented by a linear operator acting in a space of complex-valued wave functions (Hilbert space), which is defined on the configuration space of the system.

The form of an operator, in the particular case of motion of a particle in space, can be defined according to the requirement of invariance of the wave function, under the corresponding transformation of space.

2. The mathematical image of a physical system is an operator *-algebra.
The equation of motion of a particle is established for operators, which act on a wave function of a particle, such that the relation between operators is defined according to the relation between the appropriate classical physical values (the principle of correspondence).

3. A pure state of a quantum-mechanical system is characterized by an
eigenfunction  $\Psi_{i}(x)$ of a Hermitian operator $\hat{P}$, so that the expectation  of the appropriate observable  determined in the experiment
\[\hat{p}=\int_{X} \Psi^{*}(x)\hat{P}\Psi(x)dx=\sum_{i} p_{i}a_{i}a^{*}_{i}\] where $\Psi(x)=\sum_{i} a_{i}\Psi_{i}(x)$

These statements were obtained in the present theory, based on the postulate that interaction is discrete and some additional assumptions, related to this postulate. That means, the basic statements established for description of the DI-system also can be used for interpretation of the results obtained within the framework of traditional quantum theory. 

Here we may note, that the qualitative understanding, that at least in particular circumstances interaction has certain discrete features (e.g. interaction of the field is described as creation or destruction of the appropriate particles), and that parameters of the system are defined in relation to interactions, was obtained very soon after discovery of quantum mechanics.
This qualitative description, which was mostly used for interpretation of the results established within the theory, and quantitative description were not essentially connected with each other. 

However, as we see, the general statement about the discrete character of interaction implies that behavior of the system, can be described by a set of events without defined ordering. 
This suggests the specific logic of events as well as complex character of the function, which describes stochastic behavior of the system.
Analysis of the behavior of the system between interaction demonstrates that transitions of the system between the states should be described by the Hermitian operators. The expressions for the particular operators can be established using straightforward calculations and are similar to the expressions postulated within conventional quantum theory.

Modeling of behavior of the real and imaginary part of the complex function, which describes the behavior of the DI-particle, gives a way to establish the equation of motion
\\ 
(Schr\"{o}dinger) using the method similar to inferring of Smoluchowski equation in the classical statistical mechanics. The similar approach may be used to establish equations of motion for the DI-fields, which appear to have the same form as equations established within the standard quantum field theory using the principle of correspondence.
So that all known phenomenology of quantum mechanics which is based upon established mathematical formalism, in particular wave-particle duality, tunnel effects, etc., essentially follow from the model. 

\section{\rm Interpretation of particular phenomena} 
The traditionally discussed problems of interpretation of quantum mechanics, can be regarded as stemming from fundamentally positivistic character of the theory: different conceptual pictures for the same substance (wave-particle duality), coupling of the object and subject of observation (cat paradox), as well as incompatibility of the models used in the classical and quantum theories (EPR).

We will briefly discuss how the same phenomena will be interpreted within the present approach (excepting EPR, which will be considered in \cite{yudin}).
\\
\\
{\em Wave particle duality}
\\
\\
Within the present model, we have defined real particles, which demonstrate wave type behavior in space and time, as ordering of the states may be not defined in between interactions.
The principle of locality is not strictly imposed and would apply only to interactions of the specific kind.
\\
\\
{\em Interaction with the macroscopic testing device}
\\
\\
The problem of interaction with macroscopic testing device has several aspects. 
\\

(i) The problem of choosing the particular value of the parameter described by the wave function (collapse of the wave function) would not be appearing as a controversy in the theory with statistical interpretation of the wave function. Similarly, the problem will not create any kind of controversy within the present interpretation. 
\\

(ii) The problem of interfacing classical detector. The problem, widely discussed in the quantum theory of measurement, may be formulated as follows: how can the measurement be described in a way that the classical device is not in a superposed state while the tested quantum system is not in the particular defined state. 

The brief answer formulated within the framework of the present approach would sound. The DI-system is defined to stay at intermediate state as long as it does not interact with the classical external system (in between interactions); the classical system is still in its initial state. An interaction, which can change the state of the classical system, can put the DI-system into the particular defined state, which corresponds to the particular state of the classical system.

In summary, the concept of discrete interaction gives a way to de-couple the DI-system as an object of experiment and a testing device; it assumes real existence of the DI-system with attributes of substance to be always prescribed to it. This comes at expense of limitation of the principle of locality, which is considered to follow from the discrete nature of interaction. The positivistic interpretation of the phenomena may be related to attempts to expand the conventional models of space-time behavior as applied to observables, in particular the concept of ordering of states, to the state of the system in between interactions. In this case clearly several pictures may be needed to describe the same substance. In the case when the parameter of the system $\hbar\rightarrow 0$: the discrete character of interaction is changed in favor of continuos. The parameters are becoming permanently observable, so that the classical realistic description appears to be applicable.


\pagebreak
\onecolumn
\section{\rm Appendix}
The classical probability space is defined as a set 
${\cal P}\equiv\{\Omega, {\cal R}, P\}$, where $\Omega$ is a set of elementary events $a_{i}\in \Omega$, so that for $i\neq j$ $a_i\cap a_j=\emptyset$, $\cal R$ is $\sigma$-algebra of subsets of the set $\Omega$ and $P(R)$, $R\in {\cal R}$ is a probability measure, with the properties $P(\Omega)=1$, 
\begin{eqnarray}
P(\cup_{i} a_{i})=\sum_{i} P(a_{i})\nonumber\\
P(\cap_{i} a_{i})=\Pi_{i} P(a_{i})
\label{ap_prob}
\end{eqnarray}
Consider a set of pairs of events $<a_{i}|b_{j}>\in \Omega_{g}$, $i < k_{i}$, 
$j < k_{j}$ with operations of union and intersection of pairs as conventionally defined, with the property, for $i\neq j$,  $\cup_{k}^{k_j}(<a_{i}|b_{k}>\cap<b_k|a_{j}>)=\emptyset$.
We specify $\sigma$-algebra ${\cal R}_g$ of subsets of the set $\Omega_{g}$ and define g-probability space as a set 
${\cal P}_{g}\equiv\{\Omega_{g}, {\cal R}_g, \Psi \}$, where
$\Psi$ is a complex g-probability function $\Psi=\Psi(R)$, $R\in {\cal R}$ with the properties $\Psi(<b|a_{i}>)=\Psi^{*}(<a_{i}|b>)$,
$\Psi(\cup_{i} <b|a_{i}>)=\sum_{i} \Psi(<b|a_{i}>)$ and $\sum_{i,j}^{k_{i},k_{j}}\Psi(<b_{i}|a_{j}>)\Psi(<a_{j}|b_{i}>)=1$.

Note that though the direct mapping of the g-probability space ${\cal P}_g$ to probability space ${\cal P}$ is not defined, mapping of the set of pairs defined by $\Omega_g$ to the set $\Omega$ according to the formula 
\begin{equation}
a_{j}\equiv \cup_{j}^{k_i}(<a_{i}|b_{j}>\cap<b_{j}|a_{i}>),
\label{ap_el_event}
\end{equation}
so that $a_{i}\in \Omega$ would specify a classical probability space ${\cal P}$. Indeed, according (\ref{ap_el_event}) for $i\neq j$, $a_i\cap a_j=\emptyset$; by considering possible unions of events $a_i$ we define $\sigma$-algebra  $\cal R$ and consequently the probability measure $0\leq P(R)\leq 1$, 
$\forall R\in {\cal R}$. 

The probability measure $P$ is defined for the set $\{a_{i}\}$, specified by the set of pairs $<a_{i}|B>\equiv<a_{i}|b_{j}>$, $1\leq j \leq k_{j}$. 
Consider the set $\{a_i\}$ with $\sigma$-algebra, $\cal R$ and define probability measure $P(R)$ 
according to the formula
\[
P(a_{i})=\sum_{j}\Psi(<a_{i}|b_j>)\Psi(<b_j|a_{i}>),
\]
for an arbitrary subset $b_{j}\in B$, provided the relations (\ref{ap_prob}), as specified for $\cal R$. 
This defines the relation between probability $P\in {\cal P}$ and g-probability  $\Psi\in {\cal P}_g$.

\pagebreak
\onecolumn
\begin{center}
Successive transitions between the states
\end{center}
\
\\
\begin{picture}(250,230)(-60,0)
\put(10,120){1}
\put(17,123){\vector(1,0){15}}
\put(40,120){2}
\put(70,120){3}
\put(58,123){\oval(42,20)}

\put(10,50){1}
\put(27,53){\oval(42,20)}
\put(40,50){2}
\put(70,50){3}
\put(51,53){\vector(1,0){15}}

\put(170,120){1}
\put(187,123){\oval(42,20)}
\put(200,120){2}
\put(230,120){3}
\put(218,123){\oval(42,20)}

\put(170,50){1}
\put(192,53){\vector(-1,0){15}}
\put(200,50){2}
\put(230,50){3}
\put(210,53){\vector(1,0){15}}

\put(185,10){1}
\put(200,26){\vector(-1,-1){10}}
\put(200,30){2}
\put(215,10){3}
\put(205,26){\vector(1,-1){10}}
\put(202,13){\oval(42,20)}

\put(35,185){a1.}
\put(195,185){a2.}

\put(35,-8){b1.}
\put(195,-8){b2.}

\Large
\put(20,210){1}
\put(28,213){\vector(1,0){17}}
\put(50,210){3}

\put(180,210){1}
\put(197,213){\oval(50,30)}
\put(210,210){3}

\large
\put(13,-48){\vector(1,0){16}}
\put(25,-78){\oval(30,12)}
\put(50,-50){ordering defined (${\cal V}_{ij}$)}
\put(50,-80){ordering not defined 
(${\cal U}_{ij}$)}
\end{picture}
\
\\
\\
\\
\\
\\
\\
\\
\begin{center}
fig.1
\end{center}
1. For the set $\{1,2,3\}$ ordering is defined for 1 and 3.
\\
2. For the set $\{1,2,3\}$ 
no ordering is defined for 1 and 3.
\\
\\

\end{document}